\documentclass[fleqn,usenatbib]{mnras}

\usepackage{newtxtext,newtxmath}

\usepackage[T1]{fontenc}

\DeclareRobustCommand{\VAN}[3]{#2}
\let\VANthebibliography\thebibliography
\def\thebibliography{\DeclareRobustCommand{\VAN}[3]{##3}\VANthebibliography}

\usepackage{graphicx}	
\usepackage{amsmath}	
\usepackage{siunitx}
\usepackage[dvipsnames]{xcolor}
\usepackage{float}
\usepackage[normalem]{ulem}

\newcommand{\aref}[1]{\hyperref[#1]{Appendix~\ref{#1}}}

\definecolor{darkgreen}{rgb}{0.13, 0.55, 0.13}



\newcommand{\machc}{\mathcal{M}_{\rm c}}
\newcommand{\kms}{\rm km \, s^{-1}}
\newcommand{\vc}{v_{\rm c}}
\newcommand{\cs}{c_{\rm s}}
\newcommand{\va}{v_{\rm A}}
\newcommand{\mach}{\mathcal{M}}
\newcommand{\trot}{t_{\rm rot}}

\newcommand{\qeff}{Q_{\mathrm{eff}}}
\newcommand{\qgas}{Q_{\mathrm{g}}}

\defcitealias{arora_2025}{A25}

\usepackage{scalerel,tikz}
\usetikzlibrary{svg.path}
\definecolor{orcidlogocol}{HTML}{A6CE39}
\tikzset{orcidlogo/.pic={
\fill[orcidlogocol] svg{M256,128c0,70.7-57.3,128-128,128C57.3,256,0,198.7,0,128C0,57.3,57.3,0,128,0C198.7,0,256,57.3,256,128z};
\fill[white] svg{M86.3,186.2H70.9V79.1h15.4v48.4V186.2z}
svg{M108.9,79.1h41.6c39.6,0,57,28.3,57,53.6c0,27.5-21.5,53.6-56.8,53.6h-41.8V79.1z M124.3,172.4h24.5c34.9,0,42.9-26.5,42.9-39.7c0-21.5-13.7-39.7-43.7-39.7h-23.7V172.4z}
svg{M88.7,56.8c0,5.5-4.5,10.1-10.1,10.1c-5.6,0-10.1-4.6-10.1-10.1c0-5.6,4.5-10.1,10.1-10.1C84.2,46.7,88.7,51.3,88.7,56.8z};
}}
\newcommand\orcidicon[1]{\href{https://orcid.org/#1}{\mbox{\scalerel*{
\begin{tikzpicture}[yscale=-1,transform shape]
\pic{orcidlogo};
\end{tikzpicture}
}{|}}}}

\title[Magnetic destabilisation in disc galaxies]{Magnetic destabilisation in disc galaxies: Filament/feather formation}

\author[Arora et. al.]{
Raghav Arora,$^{\orcidicon{0000-0001-6427-2601}1}$\thanks{E-mail: raghav.arora@fysik.lu.se}
Oscar Agertz,$^{\orcidicon{0000-0002-4287-1088}1}$
Christoph Federrath$^{\orcidicon{0000-0002-0706-2306}2}$
and Mark R. Krumholz$^{\orcidicon{0000-0003-3893-854X}2}$
\\
$^{1}$Lund Observatory, Division of Astrophysics, Department of Physics, Lund University, Box 118, SE-221 00 Lund, Sweden\\
$^{2}$Research School of Astronomy and Astrophysics, The Australian National University, Canberra, ACT~2611, Australia
}

\date{Accepted XXX. Received YYY; in original form ZZZ}

\pubyear{\the\year{}}

\begin{document}
\label{firstpage}
\pagerange{\pageref{firstpage}--\pageref{lastpage}}
\maketitle

\begin{abstract}
Gas gravitational instability plays a crucial role in secular galactic evolution, but the role played by magnetic fields in mediating this instability -- despite their dynamical relevance and ubiquity -- is still not understood. To investigate this question we conduct a parameter study using numerical simulations of 3D isolated disc galaxies that are initialized in equilibrium, but have a range of initial magnetisation, quantified by $\beta \in \{0.1, 0.5, 1, 10, 100, \infty\}$, where $\beta$ is the ratio of thermal to magnetic pressure. We analyse how magnetic field strength influences the formation of dense filaments and feathers driven by gravitational instability. The simulations show that filament growth rate and spacing are significantly altered by dynamically strong fields ($\beta \lesssim 10$), and that these effects depend on the value of $\beta$ and strength of shear in the disc. Magnetic fields either stabilise or destabilise, with destabilisation dominating in regions with low $\beta$, and low shear. This makes the destabilisation particularly important in dwarf galaxies with low-shear rotation curves. Filament spacings are similarly affected differently in different galactic regions, depending upon the local field strength and shear. Our results are in good agreement with predictions from the magneto-Jeans mechanism. 

\end{abstract}

\begin{keywords}
galaxies: evolution -- galaxies: disc -- galaxies: magnetic fields -- galaxies: structure -- methods: numerical -- (magnetohydrodynamics) MHD 
\end{keywords}



\section{Introduction}

Gravitational instabilities are responsible for a wide range of physical processes in galaxies. They incite the formation of galaxy-wide features such as spiral arms, bars \citep{lin_shu_1964, goldreich_lynden_bell_swing_1965, julian_toomre_1966, sellwood1980_bars,bertin1989, fuchs_2001}, molecular clouds and giant gas clumps \citep{inoue_2018,dekelCeverino2009_clumps, mandelkarDekel2014, renaudRomeoAgertz_2021}, and feathers/filaments \citep{griv_wang_2014_hydro_sims_twoD, meidt_phangsjwst_2023, arora_2025}. They transport mass and angular momentum through the disc \citep{goldbaum_mass_2015, krumholzBurkhart2018}, and drive turbulence in the gas that cascades kinetic energy from hundreds of pc down to molecular cloud scales, which is a key ingredient for star formation \citep{KrumholzMcKee2005, PadoanNordlund2011, BournaudEtAl2010,HennebelleChabrier2011, FederrathKlessen2012, ejdetjarn_2022, fensch_universal_2023}. It has long been established, however, that the gravitating gas is a magnetised plasma \citep{ferriere2001_reviewISM}. Magnetic fields have been observed to thread the ISM of both the Milky Way and external galaxies \citep{beck_2015,han2017_observingMagneticFields, crutcher_review_2019, borlaff_extragalactic_2023}, at strengths that place them in rough equipartition with the turbulent and gravitational energy densities of the gas \citep{Beck2016}. Despite their dynamical significance, however, their coupling with the gas gravitational instability is still not fully understood. 

On the one hand, magnetic fields add a pressure term and therefore offer a source of stabilisation against gravity. On the other, linear instability analysis has consistently shown that they can also destabilise the disc. They do this by affecting the gas flows within the disc through magnetic stresses \citep{ balbus_hawley_1998_MRI_review}, and perpendicular to it via their buoyancy \citep{parker_1966}. The former potentially alters the balance between destabilising gravity and stabilising shear and pressure, known as the `magneto-Jeans' mechanism \citep{elmegreen_1987_magnetic, gammie_1996, kim_ostriker_2001}. The later, playing a secondary role \citep{kim_ostriker_2002}, drives gas into undulating hills and valleys that rise above and below the mid-plane of the disc via the Parker instability \citep{parker_1966, mouschovias_2009,arora_2023}. Perturbative analyses suggests that the net effect of magnetic fields can be either stabilising or destabilising, depending upon the field strength and the shape of the rotation curve \citep{ kim_ostriker_2001}.

Numerical simulations of galaxies that go into the non-linear regimes of the instability, have routinely included magnetic fields. Isolated galactic simulations have investigated magnetic field amplification and saturation from weak seed fields \citep{pakmor_simulations_2013, dobbs_pettitt_magneticReversals_2016, riederTeyssier_ssd_2016, pakmor_magnetic_2017,steinwandel2019,ntormousi2020_galacticDynamo, wissingShen2023,pakmorRebekka2024}, their effects on star formation rates \citep{whitworth_smith_2023,robinson_wadsley_2024, gurman_steinwandel_2025, ryan_rowan_whitworth_2026}, and their role in launching outflows \citep{steinwandel2019, SteinwandelEtAl2020, wibking_krumholz_2023}. Cosmological simulations that include the larger environment around the galaxy have explored their impact on disc morphologies, and the surrounding circumgalactic medium (CGM) \citep{pakmorAurigaMagnetic2017, pillepichAnnalisa2018, hopkinsFIRE2018, pakmorRebekka2024, bieriPakmor2026}. Simulations of local galactic patches benefit from an increase in spatial resolution, allowing a more physical coupling of magnetic fields with the star formation process and ISM chemistry \citep{kim_ostriker_2015,iffrig_hennebelle_2017, girichidis_2018, kim_wong_tigress_2021, brucy_2023}. The simulation setups used in these studies, however, are not suited for a systematic study of the gas gravitational instability and the interaction of magnetic field with it. While including sub-grid physics such as star formation, feedback, etc., offers increasing realism, they also make it much harder to isolate the impact of any particular phenomenon. Including sub-grid physics also increases the computational cost, which renders systematic parameter studies at a high resolution infeasible. Shearing box simulations partially offset the computational cost \citep{kim_threedimensional_2002, kim_formation_2006, lee_feathering_2014, kim_wong_kim_2015}, but they rely on WKB-like approximations, whose applicability to the global galactic context remains unclear.

There are some works in the literature describing parameter studies of magnetised isolated galaxies \citep{dobbs_price_2008, khoperskov_global_2018, inoue_2019, bastian_2019, arora_2023}. However they exclusively focus on galaxies with a fast-rising, Milky Way-like rotation curve, which can bias their findings since linear analysis shows that the magneto-Jeans instability is sensitive to the rotation profile \citep{kim_amplification_2001}. Moreover, most of these simulations do not start from controlled equilibrium conditions, which is crucial for stability analysis and comparison with linear theory. Equilibrium initial conditions have been employed in investigations of the formation of bars, spiral arms \citep{widrowEquilibriumICs2008, sellwoodGalaxyPackage2014, yurinSpringelGaelic2014, vasilievAGAMA2019, thor_nexus2024}, and recently, for kpc-sized filamentary features known as feathers \citep{arora_2025}, but thus far none of these simulations have included magnetic fields. As a result, while we know that magnetic fields will have some effect on the gas gravitational instability, a systematic investigation of this effect in 3D simulations is still lacking. 

The present study bridges this gap by carrying out controlled high-resolution 3D global simulations of isolated disc galaxies that are initialised in equilibrium, following the approach introduced in \citet[hereafter \citetalias{arora_2025}]{arora_2025}. We begin by including magnetic fields in this setup, and build a suite of simulations by varying the initial field strength. We then analyse the effects of varying the initial magnetisation on the dense feathers/filaments (hereafter referred to as filaments) that form via gravitational instability. In particular, we focus on comparing their morphology, growth rates, and spacing. We show that magnetic fields do not act as mere pressure terms. Instead, they switch between stabilising and destabilising roles, as suggested by previous analytical studies.

The paper is organized as follows: \autoref{sec:method} describes our simulation setup, initial conditions and the library of simulations used in this work. In \autoref{sec:results}, we discuss the basic morphology of our galaxies and the filaments that form in them. We further quantify the dependence of the filament growth rates and filament number/spacing on the initial magnetic field strength. In \autoref{sec:Discussion}, we compare our results with observations and existing theoretical/numerical works. Finally, we summarise our conclusions in \autoref{sec:conclusions}.

\section{Methods} \label{sec:method}
We use 3D global simulations of isolated disc galaxies. Our galaxies are isothermal, initially-axisymmetric, magnetised gaseous discs rotating with a rising rotation curve which is flat at large radii. The gaseous disc is embedded in an analytical static dark matter plus stellar potential, and is initialised in equilibrium. We outline the setup in \autoref{subsect:simulationSetup}. In \autoref{subsect:numerics} we describe the specifics of the numerics used for our simulations, and in \autoref{subsect:parameterStudy} we outline the parameter space of the simulation suite. 

\subsection{Physical Setup} \label{subsect:simulationSetup} 

Except for the addition of magnetic fields, our setup is identical to the one introduced in \citetalias{arora_2025}; the reader is referred to the work for details. Here, we briefly summarise the major physical components of the setup. 

The isothermal gas in the galactic disc is rotating with a rising rotation curve given by 
\begin{equation}\label{equation:rotationCurve}
     v_{\rm rot} = \vc\frac{R}{\sqrt{R^{2} + R_{\rm c}^{2}}}, 
\end{equation}
where $R$ is the galactocentric radius, $R_{\rm c} = 2~\rm kpc$ is the core radius of the profile, and $\vc$ is the circular velocity at large radii. The surface density of the gas in the disc follows a modified exponential radial profile that flattens close to the centre, 
\begin{equation} \label{eqn:analytical_surfaceDensity}
    \Sigma (R) = \Sigma_{\circ} \exp{\left [ -\frac{R}{R_{\rm d}} -\Gamma \exp \left (-\alpha \frac{R}{R_{\rm d}}\right ) \right]},
\end{equation}
where $\Sigma_{\circ}$ is the central surface density, $R_{\rm d} = 3~\rm kpc$ is the scale radius, and $\alpha = 2$ and $\Gamma=1/2$ are constants. The second term in square brackets serves to flatten the exponential profile near the galactic centre, as is included in order to ensure that the scale height is resolved near the centre well enough to avoid numerical artefacts due to unresolved vertical pressure gradients. The gaseous disc is immersed in a static analytical gravitational potential that accounts for the presence of both stars and dark matter. The gravitational potential approximates the stellar plus dark matter distribution by a flattened spheroid\footnote{The analytical gravitational potential used here is a logarithmic potential used to generate flat rotation profiles \citep[e.g.][]{TaskerTan2009, bastian_2019}. As a result, we treat the stellar and dark matter components as a net aggregate and do not differentiate between the two.} \citep{binney_tremaine_1987_galactic_dynamics}, which along with the gas self-gravity provides the necessary centripetal force for the rotation profile given by \autoref{eqn:analytical_surfaceDensity}. We solve for the potential using the condition for initial equilibrium. 

Note that we choose to establish initial equilibrium because we strive to study the development of gaseous instabilities in the disc. We do this by balancing all the radial and vertical force components -- including the gravitational force from the external potential, gas self-gravity, pressure, centrifugal forces in the rotating reference frame, and magnetic forces (see below) -- against each other following the method outlined by \citet{wang_equilibrium_2010}. The requirement of vertical equilibrium uniquely determines the vertical distribution of the gas at each radius, and thus this requirement together with the surface density profile and the assumption of axisymmetry fully specify the 3D gas distribution. 

The isothermal galactic disc is surrounded by a hot circumgalactic medium (CGM) with a constant temperature $T = 10^{7}~\rm K$ and a density. The density is such that there is pressure balance between the boundary of the disc and the CGM. We place the transition at an isodensity contour of $10^{-28}$ g cm$^{-3}$ in the disc, which is $\sim 5$ orders of magnitude lower than the mid-plane density. 

We initialise the magnetic field in an pre-evolved state, with a field topology that is purely toroidal throughout the simulation box. This corresponds to the dominant mode of regular magnetic fields\footnote{\textit{Regular magnetic field} component in observations refers to the magnetic field that has a well-defined direction within the beam-width of the telescope \citep{beck_chamandy_elson_2020}, which is a few hundred pc for nearby galaxies.} observed in the plane of the disc in nearby spiral galaxies \citep[][and references therein]{beck_2015}. Observations of galaxy-wide magnetic fields in nearby galaxies, however, showcase a greater complexity \citep{beck_chamandy_elson_2020}. For instance, spiral galaxies have been reported to have galaxy-wide magnetic spiral arms \citep{beck_2007_magneticArms_ngc6946, borlaff_salsaV} and field reversals \citep{han_beck_1999, giessuebel_beck_2014,beck_ic342_2015}. We neglect these sub-dominant components of the magnetic field in the initial conditions of our simulations. As our simulations evolve, we expect that the field morphology will naturally change from being purely azimuthal due to flux-freezing and field tangling. 

We scale the strength of the initial magnetic field as $ B\propto \rho^{1/2}$. This means that the magnetic pressure, given by
\begin{equation}\label{eqn:magnetic_pressure}
    P_{\mathrm{m}} = \frac{\rho \va^{2}}{2} = \frac{B^{2}}{8\pi},
\end{equation}
scales as $\propto \rho$. Here $\va = B/\sqrt{4\pi\rho}$ is the constant Alfv\'{e}n speed of the medium. Given our isothermal equation of state, this scaling ensures that the ratio of magnetic to thermal pressure is uniform throughout the disc. Such a scaling of magnetic field strength is also consistent with the results of Zeeman observations in the Milky Way, albeit only for densities $\geq 10^{3}~\rm cm^{-3}$ \citep[][and references therein]{pattel_ppvii_2023, whitworth_2025}. 

The inclusion of magnetic fields adds an additional Lorentz force term in the Euler equation \citep{shukurov_subramanian_2021} for which we must account when setting up the disc in equilibrium. We do this by approximating the Lorentz force as simply the gradient of the magnetic pressure (\autoref{eqn:magnetic_pressure}), thereby neglecting the contribution from magnetic tension. We can do so because the magnetic tension is purely radial, and its contribution in this direction is always negligible ($\leq 1\%$) in comparison to the dominant centrifugal and gravitational force terms. 

After placing the disc in equilibrium, we add turbulent velocity fluctuations in the gas in order to facilitate the development of instabilities. We do this with the publicly available code \texttt{TurbGen} \citep{federrath_2022_turbGen}. We set the amplitude of the fluctuations such that the root mean squared turbulent velocity $\sigma_v$ corresponds to a Mach number $\mach = \sigma_v/\cs = 0.5$. The velocity field has a Kolmogorov-like \citep{kolmogorov_1941} scaling of $k^{-5/3}$ on scales of $[ 50, 200 ]~\rm pc$, and contains a natural mixture of solenoidal and compressive modes \citep[see][for more details]{federrath2010_forcingInSimsObs}. 

\begin{figure}
\center
    \includegraphics[width=0.9\linewidth]{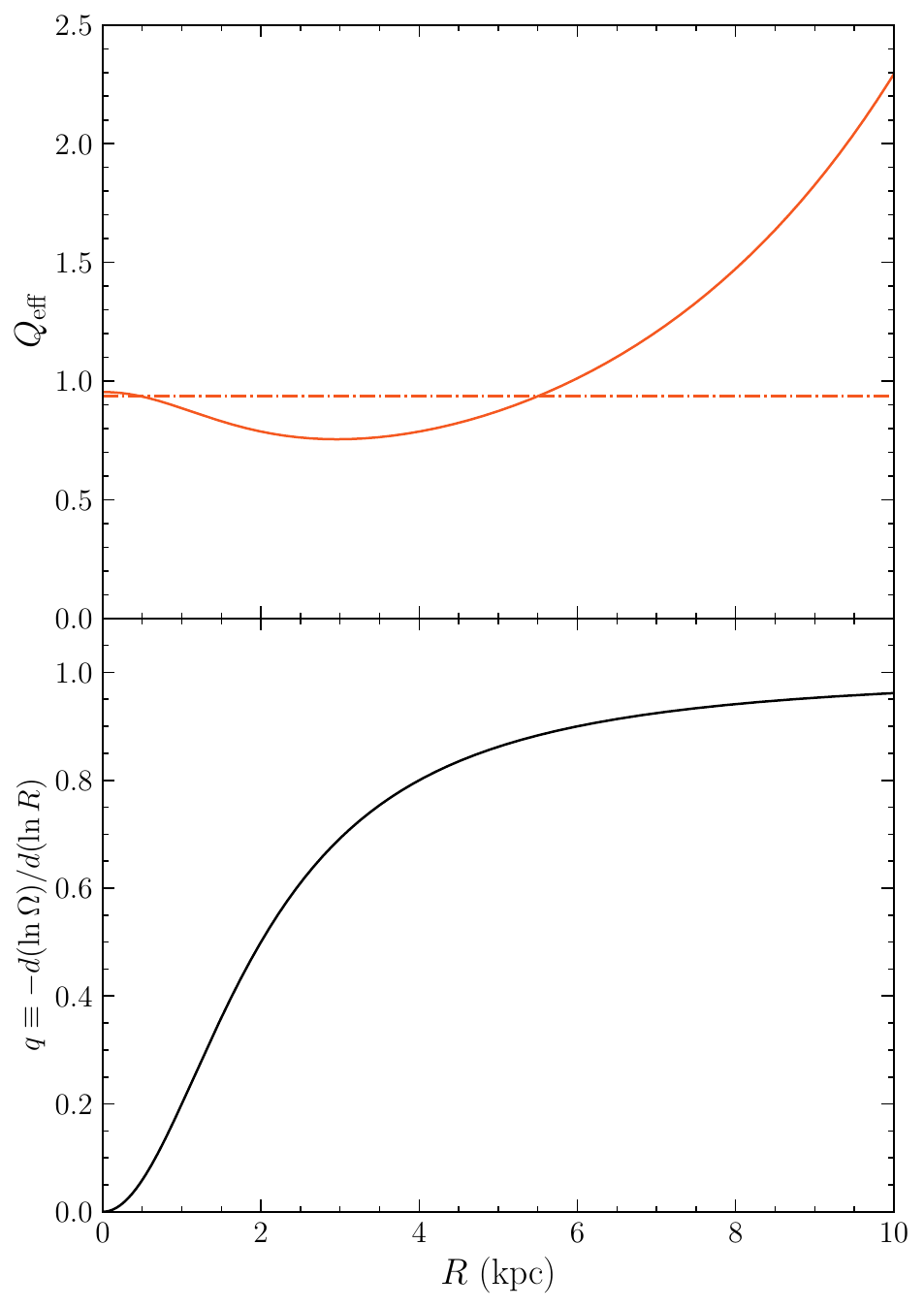}
    \caption{Top: the initial radial profile of $\qeff$ (see \autoref{eqn:q_eff}) (solid line) and its average over $0\leq R/{\rm kpc} \leq 8$ (dot-dashed line) Bottom panel: radial profile of the dimensionless shear parameter $q$.}
    \label{fig:toomre_radialProfile}
\end{figure}

\subsection{Numerics} \label{subsect:numerics}

We perform the simulations with the grid-based magnetohydrodynamical (MHD) code FLASH \citep{FryxellEtAl2000,dubey_fisher_2008} under the ideal MHD assumption \citep{choudhuri_rai_1998}. To solve the MHD equations we use the second-order accurate HLL3R Riemann scheme, which incorporates a parabolic divergence cleaning method \citep{WaaganFederrathKlingenberg2011}.

The simulation box has a length of $20~\rm kpc$ in the plane of the disc and $2.5~\rm kpc$ perpendicular to it. We use adaptive mesh refinement (AMR) to keep the region of interest uniformly resolved at a high resolution. This is a cylindrical region centred around the galactic centre, having a radius of $R_{\rm cyl} = 6~\rm kpc$ and a vertical extent of $H_{\rm cyl} = 200~\rm pc$. The base grid of the simulation has $256 \times 256 \times 32$ cells, which is uniformly increased to $4096\times 4096 \times 512$ cells in the cylindrical region. This corresponds to a base resolution of $\approx 80~\rm pc$ and a finer resolution of $\approx 4.8~\rm pc$ respectively. We resolve the region that falls outside the cylinder by a Jeans criterion, allowing the thermal Jeans length $\lambda_{\rm J} = \cs \sqrt{\pi/G \rho}$ to be resolved by a minimum of 32 cells and a maximum of 64~cells \citep{FederrathEtAl2011}.

\subsection{Parameter study} \label{subsect:parameterStudy}
The simulations are characterised by three physical dimensionless parameters:
\begin{align}
    \langle Q_{\mathrm{eff}}\rangle  & = \frac{1}{R_\mathrm{max}}\int_0^{R_\mathrm{max}} Q_{\mathrm{eff}}(R) \, dR = \frac{1}{R_\mathrm{max}}\int_0^{R_\mathrm{max}}\frac{ \kappa c_{\mathrm{eff}} }{\pi G \Sigma}\, dR, \label{eqn:q_eff} \\
    \machc & = \frac{\vc}{\cs}, \label{eq:machc} \\
    \beta & = \frac{P_{\mathrm{th}}}{P_{\mathrm{mag}}} = \frac{2\cs^{2}}{\va^{2}}, 
\end{align}
which are the radial average of the effective gaseous Toomre-$Q$, the rotational Mach number, and the plasma beta of the galaxy. Here, $R_{\rm max} = 8~\rm kpc$ is the outer boundary of the region over which we compute the average $Q_{\rm eff}$, $\kappa= \sqrt{4 \Omega^2 + 2R\Omega (d\Omega/dR)}$ is the epicyclic frequency, and $c_{\mathrm{eff}} = \sqrt{\cs^{2} + \va^{2}/2}$ is the effective sound speed of the medium accounting for both the thermal and magnetic pressure components. The Toomre-$Q$ parameter is used to characterise the gravitational stability of a razor-thin galactic disc to axisymmetric perturbations \citep{Safronov_1960, toomre_1964}. It balances the stabilising effect of rotation and pressure in the numerator with the destabilising effect of gravity in the denominator. Here, we use a modified version of Toomre-$Q$, denoted $\qeff$, in which we replace the $\cs$ by $c_{\rm eff}$ to account for the additional stabilising support due to magnetic pressure \citep{lou_zou_2006, bastian_2019}; we distinguish this from the conventional, non-magnetic Toomre $Q$ in which we omit the $\va$ term in $c_\mathrm{eff}$, which we denote $\qgas$. The circular Mach number $\machc$ quantifies the relative contribution of centrifugal to the thermal pressure support, and $\beta$ is the ratio of the thermal to magnetic pressure, which is spatially uniform for our chosen initial conditions. As shown in Appendix~B of \citetalias{arora_2025}, for a non-magnetised disc the two dimensionless parameters $\{ \langle \qgas \rangle,\machc\}$ (together with the auxiliary dimensionless shape parameters $\{R_{\mathrm{c}}/R_{\mathrm{d}}, \alpha, \Gamma\}$) uniquely determine the initial conditions. The inclusion of magnetic fields adds a third parameter $\beta$, and together this triplet of parameters $\{ \langle \qeff \rangle , \machc, \beta \}$ uniquely determine the initial conditions. 
Since we strive to quantify the effects of magnetic fields on the gas gravitational instability in a controlled experiment, in this study we select a single hydrodynamical ($\beta=\infty$) run from \citetalias{arora_2025} -- their run Q1\_$\mathcal{M}$29 -- and then repeat it while varying the strength of the initial magnetic fields to values corresponding to $\beta \in \{
100, 10, 1, 0.5, 0.1\}$. We list the dimensional parameters $\Sigma_0$ and the mean mid-plane magnetic field strength
\begin{equation}
    \langle B\rangle = \frac{\int_{0} ^{R_{\mathrm{max}}} |\vec{B}(R, z=0)|dR } {\int_{0}^{R_{\mathrm{max}}} dR}
    \label{eq:Bdefn}
\end{equation}
we derive via this approach for each of the simulations in \autoref{tab:initialConditions}.

These simulations share their initial values of $\machc = 28.4$, $\qeff$ (which is the same as $Q_{\rm g}$ for the $\beta=\infty$ run). We plot the latter in the top panel of \autoref{fig:toomre_radialProfile}. In this panel the solid line shows the initial profile of $\qeff$ while the dashed-dotted line represents its average $\langle Q_{\rm eff} \rangle = 0.94$ from $R=0$ to $R = 8\rm\,{kpc}$. In the bottom panel we show the dimensionless shear parameter $q = -d(\ln \Omega)/d(\ln R)$; for this parameter solid-body rotation corresponds to $q = 0$ and a flat rotation curve to $q = 1$. \citetalias{arora_2025} found that their simulation with this $\qeff$, $q$ profile is gravitationally unstable to filaments and forms them within a single rotation. This makes it an ideal candidate for the current study. The choice of $\machc=28.6$ for these runs is motivated by its proximity to the median value observed in nearby galaxies \citep{lang_meidt_2020} (assuming a sound speed $\cs = 7~\rm km~s^{-1}$, typical for the warm neutral medium of galaxies -- \citealt{naomi_2023_hi}). 

Note that our setup of fixing $\qeff$ rather than $\qgas$ for the purpose of studying magnetic fields is not universal in the literature; some authors instead choose to fix $\qgas$. We discuss the effects of his choice in \aref{appendix:runsWithSameQg} and \autoref{subsubsect:comparisonWithSimulations}.

\begin{table}
\centering
\caption{Input and derived parameters for the simulations. Here $\langle B\rangle $ is the mean initial magnetic field strength (\autoref{eq:Bdefn}) and core surface density $\Sigma_\circ$ (\autoref{eqn:analytical_surfaceDensity}). The galaxies share their $\qeff$ profile, with mean value $\langle \qeff \rangle = 0.94$ (c.f.~\autoref{fig:toomre_radialProfile}), and $\machc = 28.6$. We set the dimensional circular speed and sound speed to $\vc = 200~\rm \kms$ and  $\cs = 7~\kms$, respectively.} 
\begin{tabular}{ccccccc} 
\hline 
\hline
 Model   & $\beta$   & $\Sigma_{\circ}$ & $\langle B \rangle$ \\
  Name & & $(\rm M_{\odot} \, pc^{-2})$  & $(\mu \rm G)$\\
\hline
$\beta=\infty$             &  $\infty$   &  179 & 0 \\
$\beta=100$              &  100   &  180 & 1.8 \\
$\beta=10$            &   10    &  188 & 5.6 \\
$\beta=1$                &  1   &  253 & 17.7 \\
$\beta=0.5$               &  0.5   &  310 & 26.5 \\
$\beta=0.1$               &  0.1   &  594 & 55.9 \\
\hline
\hline
\label{tab:initialConditions}
\end{tabular}
\end{table}

\begin{figure*}
    \centering
     \includegraphics[width=0.95\linewidth]{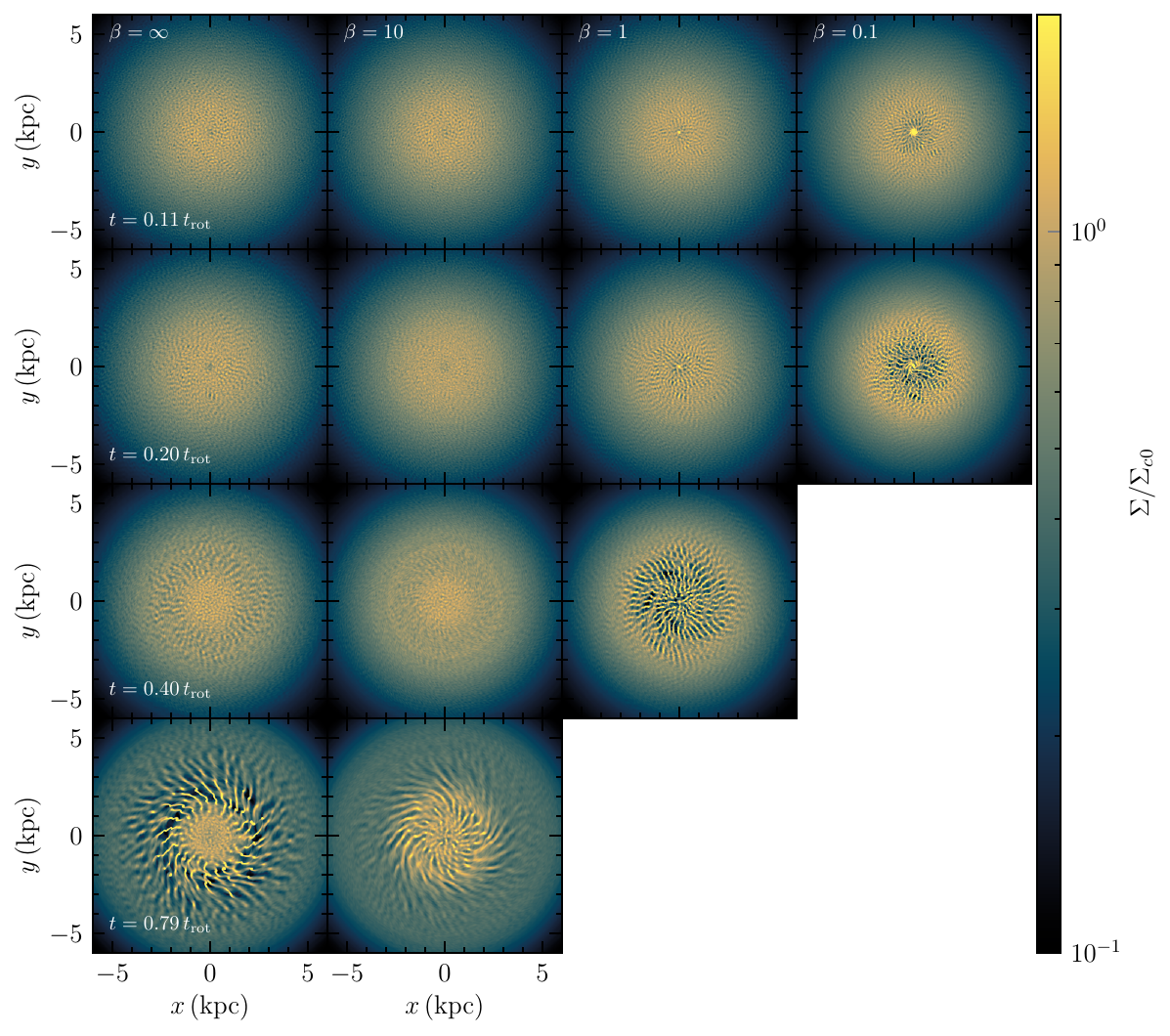}
    \caption{Each panel shows the projected density of the runs onto the $z$-plane, normalised with the central projected density $\Sigma_{c0} = \Sigma(R=0)$ of the respective run. The columns from left to right are runs with the same initial $\langle \qeff \rangle$ but with decreasing (increasing) initial $\beta$ (magnetisation). The rows from top to bottom show the time evolution of the galaxies, as indicated in the legend. The time labels depict the time in units of the galactic rotation period calculated at $R = 3~\rm kpc$. The runs are shown till the time we can reliably resolve the dense filaments with the current resolution. We can see that the addition of magnetic fields moves the region of filament formation radially inwards. Compared with the $\beta = \infty$ run, the $\beta = 10$ case forms more diffusive and thicker filaments for identical times. With increasing magnetisation, for $\beta\leq 1$, these features emerge faster, and are thinner and more finely spaced.}
    \label{fig:projection_evolution}
\end{figure*}

\section{Results} \label{sec:results}
\subsection{Disc morphology} \label{subsect:discMorphology}
In \autoref{fig:projection_evolution} we show the face-on surface densities of a subset of the simulations listed in \autoref{tab:initialConditions}; columns show different simulations (from $\beta=\infty$ to $\beta = 0.1$, left to right), while rows show different times within a single simulation. We omit the $\beta = 100$ and $\beta=0.5$ runs because their morphological evolution is similar to the $\beta =\infty$ and $\beta=1$ cases, respectively. Times in the different rows are expressed in units of the rotation period of the galaxy, given by $\trot = 2\pi R/v_{\mathrm{rot}}(R)$, at $R = 3~\rm kpc$. We halt our runs at the point where $>5\%$ of the mass in the domain is in cells that violate the \citet{TrueloveEtAl1997} criterion with a Jeans number $J=4$ (i.e., the cell spacing $\Delta x < J^{-1} \cs \sqrt{\pi/G \rho_{\rm gas}}$), since beyond this point the evolution is dominated by numerical fragmentation and becomes unreliable; blank panels in \autoref{fig:projection_evolution} corresponds to times beyond where we halt the runs due to this condition.

Looking at the $\beta = \infty$ case in the first column, we see that there are some minor density fluctuations present in the disc at $t = 0.11~\trot$. These are due to the initial turbulent velocity fluctuations we introduce at the start of our simulation. These initial fluctuations develop into faint filaments by $t\sim 0.4~\rm\trot$, and by $0.8~\rm \trot$ they transform into distinct filaments with some developing circular over-densities. The formed filaments are dense, kpc-sized, and are regularly spaced throughout the azimuth, and as seen in \citetalias{arora_2025} they form solely in the presence of gravity, pressure, and rotation. Moreover, they are absent in the region $R \leq 1.5~\rm kpc$, likely due to the greater stability of this region in our initial conditions -- as shown in \autoref{fig:toomre_radialProfile}, the inner regions of the galaxy have both a higher $\qeff$, inhibiting the growth of axisymmetric instabilities \citep[e.g.][]{toomre_1964}, and lower shear ($q$), inhibiting the growth of non-axisymmetric instabilities such as Swing amplification \citep[e.g.][]{goldreich_lynden_bell_swing_1965, julian_toomre_1966, jog_1992_swing, fuchs_2001,wang_equilibrium_2010, binney_2020,meidt_2024}. 

Moving over to the magnetic field runs, we see that the evolution of the $\beta = 10$ run is similar to the $\beta = \infty$ case till $t = 0.40~\trot$. However, by $0.8~\trot$, we see considerable differences. First, the region of filament formation has moved inwards to $R\leq 1.5 ~\rm kpc$. Second, at similar evolution time the resultant filaments have a lower density contrast and are thicker than the $\beta = \infty$ run. For equipartition magnetic field strengths and higher (i.e.~$\beta \leq1$), we see a significant change in the evolution of the galaxy even at very early times. The dense filaments in this case are thinner, more finely spaced, and emerge progressively faster with decreasing $\beta$. Filaments in the $\beta = 1$ run reach the same density contrasts by $t = 0.4~\trot$ that the $\beta = \infty$ run achieves only at $t = 0.8~\trot$. For $\beta = 0.1$, this density contrast is reached even earlier, at $t = 0.20~\trot$. Moreover, similar to the $\beta = 10$ run, these cases with a higher magnetisation also lack the inner radial cut-off of filament formation observed in the $\beta = \infty$ run. This suggests that magnetic fields alter the nature of gravitational instability responsible for filament formation.

\subsection{Filament formation rates} \label{subsect:filamentFormation_timescales}
\begin{figure*}
    \centering
     \includegraphics[width=0.99\linewidth]{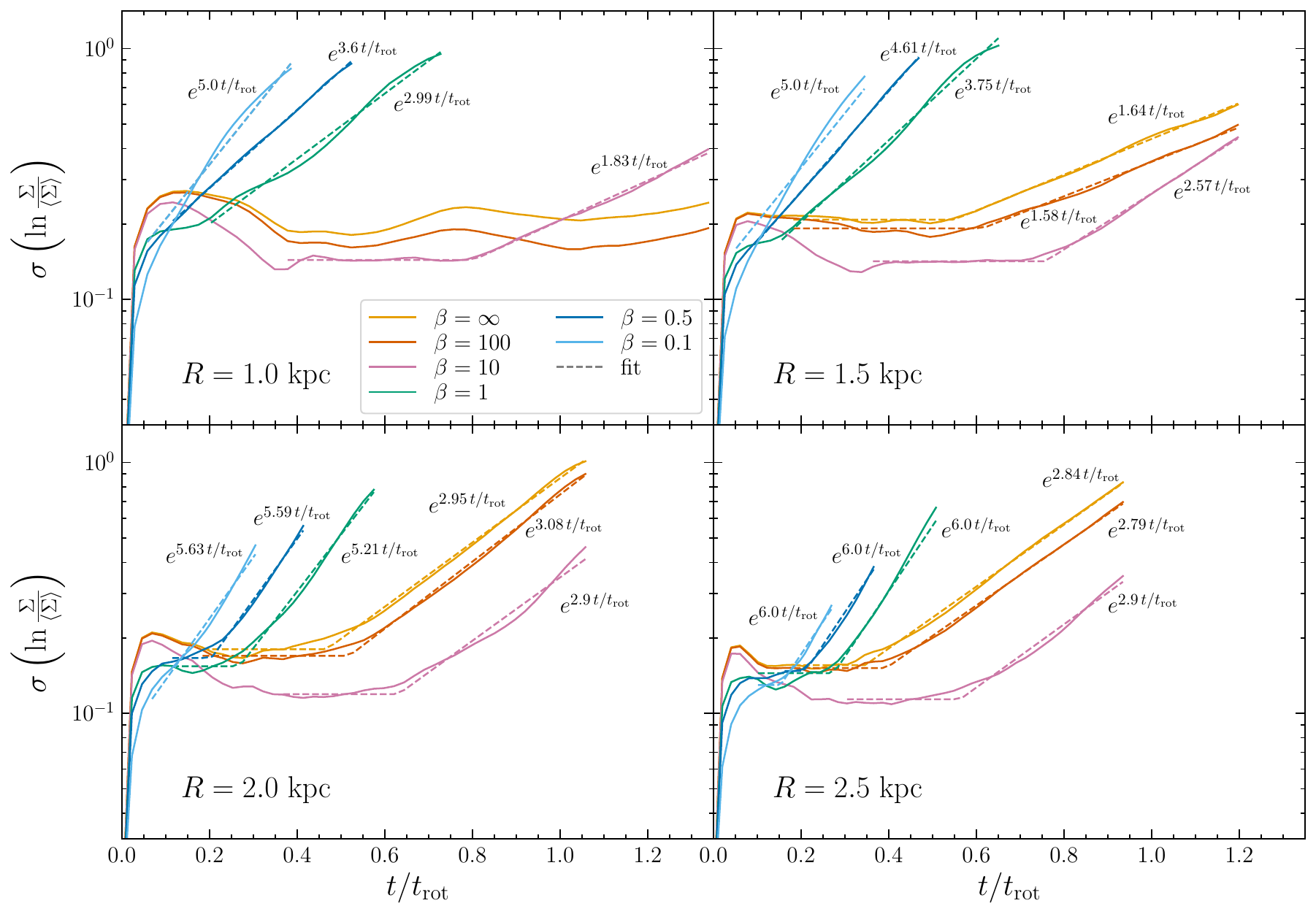}
    \caption{ The time evolution of the standard deviation of logarithmic surface density for the simulations. The different panels show the same quantity at different galactocentric radial bins as indicated in the legend, where each bin has a thickness of $100~\rm pc$. We normalise the time on the horizontal axis with $\trot = 2 \pi \vc/R$, where $\trot$ is the rotation period at $R$, which is taken to be the centre of the radial bin. The solid lines in each panel show the data from the simulation, and the dashed lines are piecewise empirical constant + exponential fits (see \autoref{subsect:filamentFormation_timescales} for details). The exponential fit parameters are shown next to the curves. Comparing with \autoref{fig:projection_evolution}, we see that the presence (absence) of the exponential rise correlates with filament formation (its lack), and that for curves that do rise, the exponent of the rise varies with $\beta$ and $R$. 
    }
    \label{fig:timeEvolotion_standardDeviation_relativeSurfaceDensity}
\end{figure*}

Here we quantify the effects of magnetic fields on filament formation rates. We do this by examining radially-binned distributions of $\ln (\Sigma/\langle \Sigma\rangle)$, following the procedure outlined in \citetalias{arora_2025}. \citetalias{arora_2025} found that the standard deviation of the $\ln (\Sigma/\langle \Sigma\rangle)$ distribution rises exponentially at radii where filament formation occurs. Physically, this is due to the coupled emergence of over-dense filaments and under-dense regions between them, which widens the distribution of $\ln (\Sigma/\langle \Sigma\rangle)$. We show $\sigma \left ( \ln (\Sigma/\langle \Sigma\rangle) \right )$ as a function of time at galactocentric radii $R = \{1.0, 1.5, 2.0, 2.5\}~\rm kpc$ (corresponding to the radii at which we see filament formation) in \autoref{fig:timeEvolotion_standardDeviation_relativeSurfaceDensity}; the result shown for each radius is the standard deviation computed over a bin $100 \, \rm pc$ thick centred on that radius. Solid lines are data from the simulations and the dashed lines are empirical fits to the data, which are described at the end of this section. 

At $R = 1~\rm kpc$ (top-left panel), we see that all the runs show an initial jump in $\sigma$ at the start of the simulation. As noted in \citetalias{arora_2025}, this happens despite the initial conditions that are analytically in equilibrium, because there is transient adjustment due to the finite resolution of the numerical grid, and to the initial turbulent velocity fluctuations we introduce at $t = 0$ (see \autoref{subsect:numerics}). The evolutionary paths of the curves after the initial transient phase, however, diverge. The $\beta = 10, 1, 0.5,$ and $0.1$ runs all exhibit an exponential rise in $\sigma$. Looking at \autoref{fig:projection_evolution}, we see that this exponential rise corresponds to the presence of filaments at $R=1~\rm kpc$. The $\beta = \infty$ and $100$ cases on the other hand show a slight temporary decrease in $\sigma$ after the initial transient phase, and then settle onto a plateau by $\sim 0.4~\trot$. This stagnant $\sigma$ correlates with the absence of filaments at $R = 1~\rm kpc$ for the $\beta = \infty$ and $100$ runs. For the $\beta \leq 10$ runs where filaments do form, we also see that the slope of the exponential rise steepens with decreasing $\beta$, reflecting more rapid filament formation.

\begin{table}
\setlength{\tabcolsep}{3.5pt} 
\caption{Fit parameters for the time evolution of $\sigma \left ( \ln(\Sigma/\langle \Sigma \rangle) \right )$. The empirical fit is performed using the functional form given by \autoref{eqn: piece-wise_fit}. We do not fit the $\beta = 100$ and $\beta = \infty$ cases at $R = 1~\rm kpc$ because they do not exhibit growth. Blank $t_{\rm onset}$ values indicate runs where we fit a pure exponential function because no plateau stage is visible. The best fits are shown as dashed lines in \autoref{fig:timeEvolotion_standardDeviation_relativeSurfaceDensity}.} \label{table:piece-wise fit}
\centering
\begin{tabular}{c c c c c c} 
\hline \hline
Model Name & $R~(\rm kpc)$ & $A$  & $\omega \trot$        & $t_{\rm onset}/t_{\rm rot}$  & $t_{\rm relax}$  \\
\hline
$\rm \beta=\infty$  &  1.0  & - & - & - & - \\
    & 1.5 & $0.08 ^{+0.001} _{-0.001}$ & $1.65 ^{+0.02} _{-0.01}$  & $0.55 ^{+0.01} _{-0.01}$ & $0.18$ \\
    & 2.0 & $0.04 ^{+0.003} _{-0.004}$ & $2.96 ^{+0.11} _{-0.08}$  & $0.47 ^{+0.02} _{-0.02}$ & 
    $0.18$\\
    & 2.5 & $0.06 ^{+0.002} _{-0.003}$ & $2.84 ^{+0.07} _{-0.06}$ & $0.35 ^{+0.01} _{-0.01}$ & 
    $0.18$\\
\\
$\beta=100$   & 1.0  & -  & - & - & - \\
    & 1.5 & $0.07 ^{+0.002} _{-0.003}$ & $1.59 ^{+0.04} _{-0.04}$ & $0.61 ^{+0.02} _{-0.01}$ & $0.18$ \\
    & 2.0 & $0.03 ^{+0.003} _{-0.002}$ & $3.07 ^{+0.08} _{-0.11}$ & $0.52 ^{+0.01} _{-0.02}$ & 
    $0.18$\\
    & 2.5 & $0.05 ^{+0.002} _{-0.002}$ & $2.79 ^{+0.06} _{-0.06}$ & $0.39 ^{+0.01} _{-0.01}$ & 
    $0.18$\\
\\
$\beta=10$   & 1.0  &$0.03 ^{+0.002} _{-0.001}$ & $1.83 ^{+0.04} _{-0.06}$ & $0.80 ^{+0.01} _{-0.01}$ & $0.35$  \\
    & 1.5 & $0.02 ^{+0.002} _{-0.002}$ & $2.58 ^{+0.07} _{-0.07}$ & $0.76 ^{+0.01} _{-0.01}$ & $0.35$ \\
    & 2.0 &$0.02^{+0.004} _{-0.001}$  & $2.90 ^{+0.06} _{-0.29}$& $0.63 ^{+0.01} _{-0.02}$ & $0.35$\\
    & 2.5 & $0.02 ^{+0.003} _{-0.004}$ & $2.90 ^{+0.22} _{-0.19}$& $0.56 ^{+0.02} _{-0.01}$ & $0.30$ \\
\\
$\beta=1$   & 1.0 & $0.13^{+0.005} _{-0.005}$  & $2.71 ^{+0.08} _{-0.08}$  & - & $0.20$ \\
    & 1.5 & $0.10^{+0.003} _{-0.003}$ & $3.75 ^{+0.06} _{-0.06}$  & - &  $0.15$ \\
    & 2.0 & $0.04 ^{+0.005} _{-0.006}$ & $5.22 ^{+0.33} _{-0.26}$ & $0.27 ^{+0.02} _{-0.01}$ & $0.10$ \\
    & 2.5 & $0.03^{+0.001} _{-0.001}$ & $6.00 ^{+0.01} _{-0.01}$ & $0.28 ^{+0.01} _{-0.01}$ & $0.10$\\
\\
$\beta=0.5$   & 1.0 & $0.14^{+0.001}_{-0.001}$  & $3.60^{+0.03}_{-0.03}$  & - & $0.10$ \\
& 1.5 & $0.11^{+0.001}_{-0.001}$ & $4.61 ^{+0.02}_{-0.02}$  & - & $0.10$ \\
& 2.0 & $0.05 ^{+0.009}_{-0.007}$ & $5.62^{+0.55}_{-0.37}$ & $0.20^{+0.01}_{-0.01}$ & $0.10$\\
& 2.5 & $0.04^{+0.007}_{-0.001}$ & $6.00^{+0.01}_{-0.73}$ & $0.21^{+0.01}_{-0.01}$ & $0.15$ \\
\\
$\beta=0.1$   & 1.0 & $0.13^{+0.008}_{-0.008}$ & $5.00 ^{+0.02} _{-0.02}$ & - & $0.05$ \\
    & 1.5 & $0.12^{+0.011}_{-0.011}$ & $5.00 ^{+0.03} _{-0.03}$&  - & $0.05$\\
    & 2.0 & $0.08^{+0.003} _{-0.003}$ &$5.63 ^{+0.17} _{-0.17}$ & - &  $0.05$\\
    & 2.5 & $0.05^{+0.009}_{-0.001}$ & $6.00 ^{+0.01} _{-0.89}$& $0.15^{+0.01} _{-0.01}$ & $0.10$\\
\hline
\end{tabular}
\end{table}


To quantify the rate of filament growth, we fit the curves shown in \autoref{fig:timeEvolotion_standardDeviation_relativeSurfaceDensity} to a functional form 
\begin{equation} \label{eqn: piece-wise_fit}
    \sigma(t) = \begin{cases}
     A e^{\omega t_{\rm onset}} & t\leq t_{\rm onset}, \\
    A e^{\omega t} & t > t_{\rm onset},
    \end{cases}
\end{equation}
where the first half of the function is a constant, representing the plateau $\sigma$, and the second half is an exponential with an amplitude $A$ and the exponent $\omega$; the two parts are separated via the onset time $t_{\rm onset}$. We fit this functional form to the data using $\chi^2$ minimisation, and excluding times $t < t_\mathrm{relax}$, where $t_\mathrm{relax}$ is the duration of the initial transient at the simulation start; we estimate $t_\mathrm{relax}$ by eye, and report the values we use below.
We also note that for some cases, such as the $\beta = 0.1$ run at $R =1 ~\rm kpc$, the plateau phase is absent, and for these we fit a pure exponential profile starting at $t = t_{\rm relax}$. We also forgo fitting entirely for the cases that do not have an exponentially rising part, such as the $\beta = 100$ run at $R = 1~\rm kpc$. We calculate the errors in the fitted parameters using jackknife resampling. We carry out $1000$ fits, each on a randomly-selected sub-set amounting to $40\%$ of the whole data. We take the 16th and 84th percentile values of the best-fit parameters produced by this procedure as our 68\% confidence interval.

\begin{figure}
    \centering
     \includegraphics[width=0.99\linewidth]{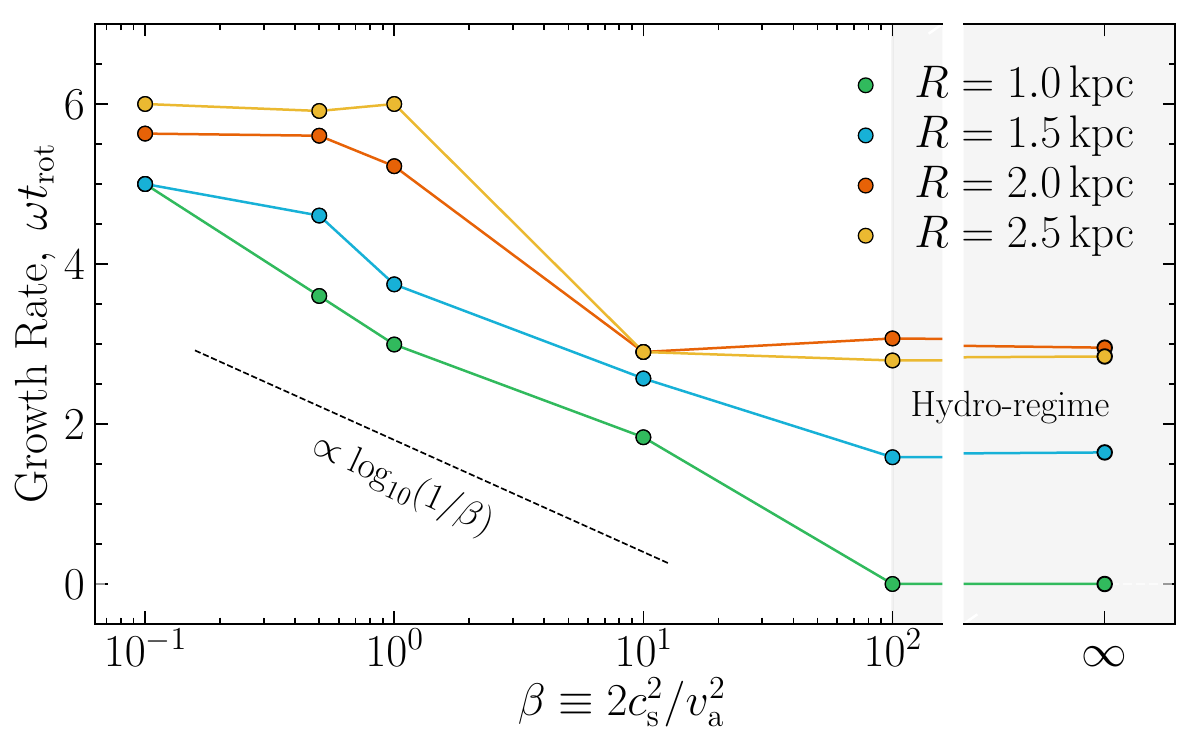}
    \caption{Dimensionless growth rate (see \autoref{table:piece-wise fit}) of filament formation as a function of the initial $\beta$ for different galactocentric radii $R$, corresponding to the different panels in \autoref{fig:timeEvolotion_standardDeviation_relativeSurfaceDensity}. The dashed line shows a $\propto \log_{10}(1/\beta)$ scaling to guide the eye. This scaling approximately matches the measurements for $\beta \leq 100$ for the galactic region at $R = 1.0, 1.5~\rm kpc$, and in the range $\beta \in (1, 10)$ for $R = 2.0, 2.5~\rm kpc$. 
    }
    \label{fig:growthRate_with_pbeta}
\end{figure}

We report the best-fit parameters (computed on the full data set) and their uncertainties in \autoref{table:piece-wise fit}, and plot the fits as dashed lines in \autoref{fig:timeEvolotion_standardDeviation_relativeSurfaceDensity}. We see that the fits trace the simulation data reasonably well. We plot the dimensionless growth rate $\omega \trot$ for all the galaxies considered in \autoref{fig:timeEvolotion_standardDeviation_relativeSurfaceDensity} against the initial $\beta$ of the runs in \autoref{fig:growthRate_with_pbeta}. The different colours correspond to different radial bins, identical to the ones used in \autoref{fig:timeEvolotion_standardDeviation_relativeSurfaceDensity}.



We see a clear rise in the filament formation rate with decreasing (increasing) $\beta$ (magnetic field strength) at all galactocentric radii. However, this increase in the growth rate only starts below (above) a critical threshold plasma beta (magnetic field strength), which we denote as $\beta_c$. While for $\beta < \beta_c$, the growth rate increases with decreasing $\beta$, for $\beta \geq \beta_c$, magnetic fields have a negligible effect on the growth rate. One can get a rough estimate of $\beta_c$ from the figure. For the inner parts of the galaxy with $R = 1.0, 1.5~\rm kpc$, $\beta_c$ lies in the range $(10, 100]$. Further out at $R = 2.0, 2.5~\rm kpc$, we see that it shifts towards a lower value in the range $(1, 10]$. This connects back to what we saw in \autoref{fig:projection_evolution}, where magnetic fields induced filament formation in the region $R \leq 1.5~\rm kpc$ for $\beta\leq 10$, while in the region $R\geq 2.0~\rm kpc$ magnetic fields increased filament formation rates starting at $\beta \leq 10$.

A number of notable differences in the inner and outer galaxy remain; for the inner two radii, we see that for $\beta \leq \beta_c$, the growth rate continuously rises with a linear dependence upon $\log_{10}(1/\beta)$. For the outer two radii, we see that the growth rate also rises, but saturates at $\beta \sim 1$. With the coarse sampling of the parameter space available, we cannot discern the exact nature of this rise. Nonetheless, we see that the shift in $\beta_c$ to lower values, and the appearance of the growth rate plateau as we move radially outwards, effectively pinches the range of $\beta$ for which magnetic fields destabilise and increase the growth rate of filaments, limiting the destabilisation range from $\beta < 100$ for $R = 1.0, 1.5~\rm kpc$ to $\beta \in (1, 10)$ at $R = 2.0, 2.5~\rm kpc$. In \autoref{subsect:natureOfInstability} we discuss physical reasons for the increase in growth rates with decreasing $\beta$ and the radius-dependent nature of this increase. 

Finally we can examine the growth rate as a function of galactocentric radius at fixed $\beta$. For the $\beta = \infty$ case it is clear that the growth rate increases as we move radially outwards, which as we discuss further in \autoref{subsect:discMorphology} could either be due to a decrease in $\qeff$, or an increase in the shear ($q$ value) as we move radially out from $R = 1~\rm kpc$ to $R = 2.5~\rm kpc$ (see \autoref{fig:toomre_radialProfile}). Both increase the susceptibility of the disc to gravitational destabilisation, and thus increasing the filament formation rates. We see the same trend for lower $\beta$ values. The difference between the growth rates at different radii, however, decreases by a factor of $3$ for the $\beta = 10, 0.1$ runs. This is due to the two growth rate plateaus seen at $\beta\geq 10$ and $\beta \leq 1$ for $R = 2.0, 2.5 ~\rm kpc$ as noted above.

\subsection{Filament spacing} \label{subsect:filamentspacing}

\begin{figure*}
    \centering
     \includegraphics[width=0.95\linewidth]{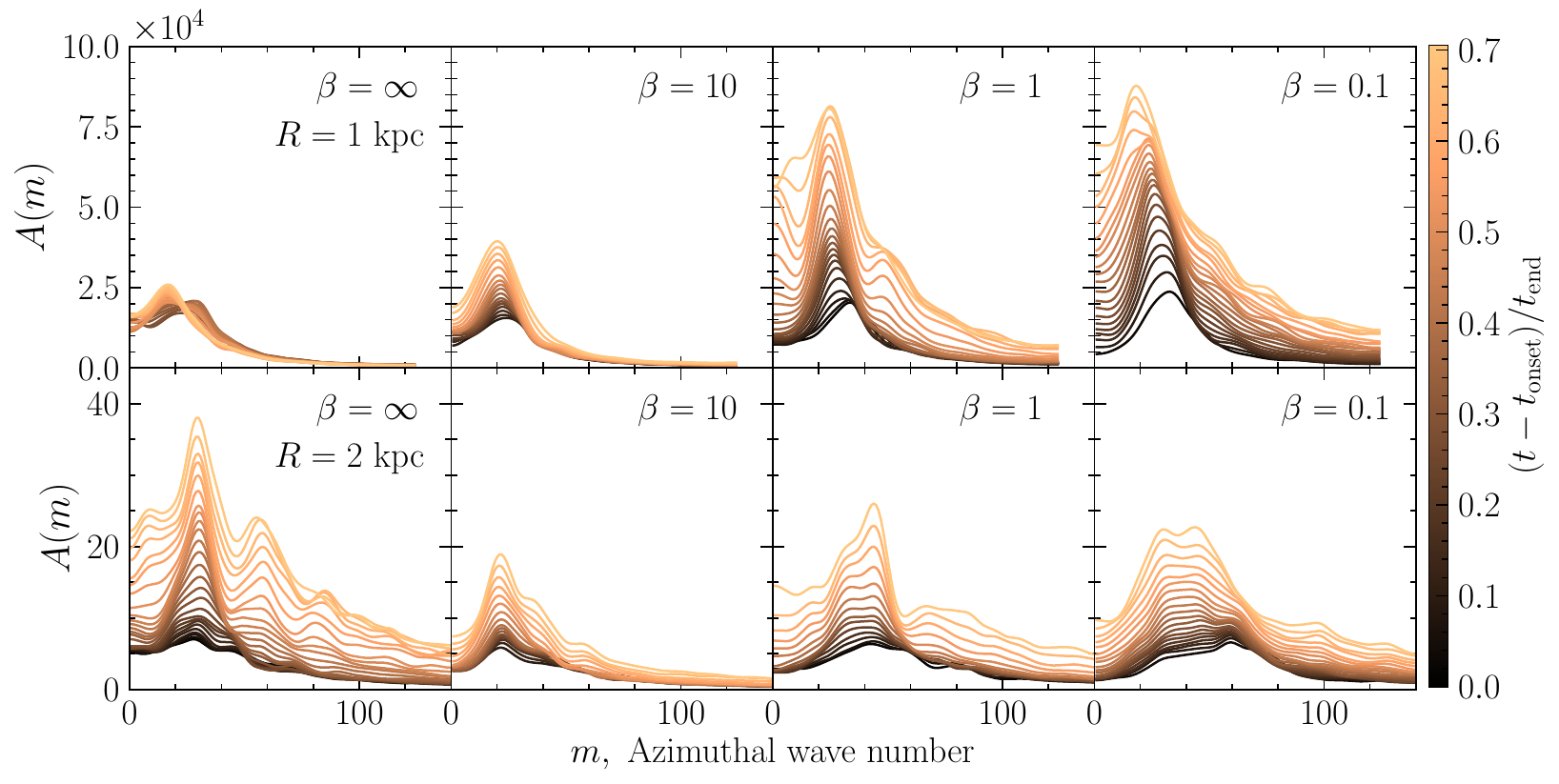}
    \caption{The azimuthal Fourier amplitude $A(m)$ as a function of the azimuthal wave number $m$ for the same set of runs as in \autoref{fig:timeEvolotion_standardDeviation_relativeSurfaceDensity}. The different columns show the amplitude for runs with decreasing $\beta$ from left to right. The first row and the second row show the Fourier amplitude in $100\rm~pc$-wide bins centred at $R=1~\rm kpc$ and $R = 2~\rm kpc$, respectively. The different curves in each panel corresponds to a particular time snapshot in the window of filament formation, $t\in \{t_{\mathrm{onset}}, t_{\mathrm{end}} \}$, as indicated by the colourbar. Except for the $\beta = \infty$ case at $R = 1~\rm kpc$, we see that curves for all $\beta$ values have a distinct peak that rises with time. The rise reflects the emergence of uniformly spaced filaments in the simulations. The amplitude of the peak is proportional to their density contrast at the end of the simulation.}
    \label{fig:a_m_with_time}
\end{figure*}

As seen in \autoref{subsect:discMorphology}, magnetic fields affect the azimuthal filament spacing of our galaxies. Here, we quantify this effect. We use the method outlined in \citetalias{arora_2025}, wherein we capture the regularity of filaments in our galaxies via Fourier transform in the $(\ln R, \theta)$ coordinate space. Here, $R$ is the radius in kpc and $\theta$ is the azimuthal coordinate. We compute
\begin{equation}\label{eqn:2d_fourier_transform}
    A(m,p) = \int ^{u_{\mathrm{max}}} _{u_{\mathrm{min}}} \int ^{2\pi} _{0}  \ln \left (\frac{\Sigma}{\langle \Sigma \rangle} \right ) e^{i(m\theta + pu)} \, d\theta\, du,
\end{equation}
where $A(m,p)$ is the 2D amplitude of the Fourier transform, $u = \ln R$ and $(u_{\mathrm{min}}, u_{\mathrm{max}})$ define the bounding radial coordinates of the region of interest in our galaxy, and $m$ and $p$ are the azimuthal and radial wavenumber, respectively. This is equivalent to a decomposition of the relative surface density in terms of a family of logarithmic spirals. The spirals are geometrically characterised by the pair $(m,p)$, where $m$ is the number of spirals spaced at a constant azimuthal spacing of $2\pi R/m$, with a pitch angle $\tan^{-1}(-m/p)$. Next, we integrate the 2D Fourier amplitude over the radial wavenumber to get the azimuthal power $A(m) = \int A(m,p)dp$, which is independent of the pitch angle. Thus, peaks in the function $A(m)$ indicate the presence of features with a high density contrast that have an $m$-fold symmetry along the azimuth. 

We plot the amplitude $A(m)$ against the azimuthal wave number $m$ of our runs in \autoref{fig:a_m_with_time}. The two rows in the figure correspond to $100~\rm pc$ wide radial bins centred at $R = \{1.0, 2.0\}~\rm kpc$, and the different columns show the runs with varying initial magnetisation. Individual curves in each panel are coloured to illustrate the time evolution of $A(m)$, as indicated by the colourbar. To ease comparison across the runs, which evolve at different rates (as seen in \autoref{subsect:filamentFormation_timescales}), we colour the curves with an effective dimensionless time. We do this because the simulations are stopped at varying end times, which is the time at which a significant fraction ($\gtrsim 10\%$) of cells violate the Truelove criterion \citep{TrueloveEtAl1997}. We define the effective time as the time relative to the onset time of filament formation $t_{\rm onset}$ normalised with the end time of the respective simulation $t_{\rm end}$. For the simulations that do not have a well-defined $t_{\rm onset}$, such as the ones that do not form filaments or the ones where filament formation starts close to $t = 0$ (see \autoref{subsect:filamentFormation_timescales} for details), we use $t_{\rm onset} = 0.1~\trot$ as a fiducial value. Finally, to smooth-out fluctuations in $A(m)$ due to the discrete numerical grid, we convolve each curve with a 1D Gaussian kernel having a standard deviation of $m = 4$, which aids in identifying the peak of the distributions.

The amplitudes $A(m)$ at $R = 1~\rm kpc$ (top row of \autoref{fig:a_m_with_time}) show a smooth rise in power that peaks at a distinct $m$ for all runs. This distinct peak indicates the presence of filaments that have a high density contrast and are spaced regularly in azimuth. 
For the $\beta = \infty$ case, we see that the amplitude of $A(m)$ does not show appreciable change with respect to its initial value, corresponding to the absence of filaments for this run at $R = 1~\rm kpc$. In contrast, the $\beta = 10$ case displays a continuous rise in $A(m)$, while maintaining the location of its peak. This corresponds to the appearance of filaments in the inner $1~\rm kpc$ with magnetic fields. As expected from increasing magnetic destabilisation, the peak in the $\beta = 1, 0.1$ runs increases to values that are a factor of $2$ higher than the $\beta = 10$ case. 

By contrast at $R = 2~\mathrm{kpc}$,
the $\beta = \infty$ curves show a consistent rise with time, reflecting filament formation of the hydrodynamical case at larger radii, while the $\beta = 10$ run has a lower amplitude than the $\beta = \infty$ one, consistent with the more diffuse filaments seen in this case for comparable times. We note that this is not due to smaller growth rates of the instability, which differ only slightly at this radius (blue curve in \autoref{fig:growthRate_with_pbeta}), but instead reflect that the instability is weaker at comparable times for $\beta = 10$ due to its lower starting amplitude and longer plateau phase (\autoref{fig:timeEvolotion_standardDeviation_relativeSurfaceDensity}). As we move on to the $\beta = 1, 0.1$ cases, we see that the peak shifts to a higher $m$-value with an increase in the peak's spread in the $\beta = 0.1$ case. The increased spread is a natural consequence of the increase in the complexity of the formed filaments at $\beta = 0.1$.

\subsubsection{Comparison with linear theory} \label{subsect:comparisonWithLinearTheory}

\begin{figure}
    \centering
     \includegraphics[width=0.95\linewidth]{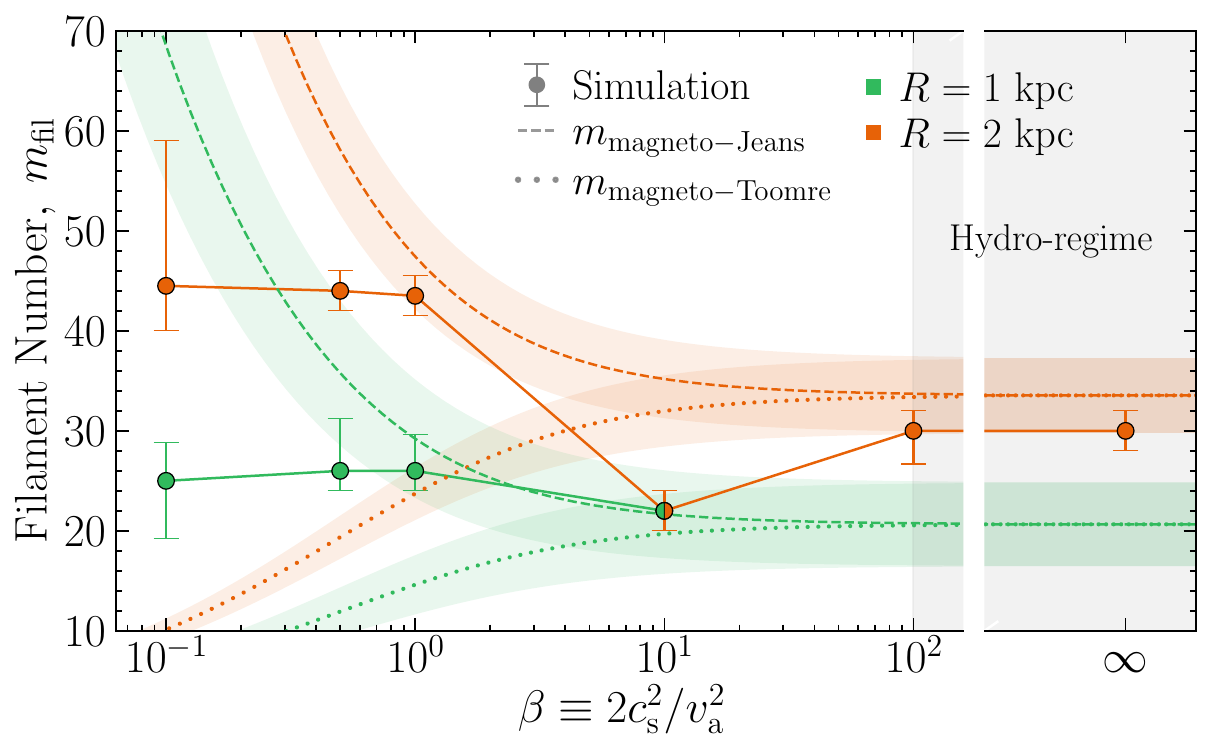}
    \caption{The number of filaments, $m_{\rm fil}$, with respect to the initial $\beta$. Circular points are data from the simulations, where the different colours correspond to the two galactocentric radial bins centred at $R = 1,2~\rm kpc$, similar to \autoref{fig:a_m_with_time}. The circular points show the median location of the peak of $A(m)$ (see \autoref{fig:a_m_with_time}) in the time-window of filament formation, given by $t \in (t_{\mathrm{onset}}, t_{\mathrm{end}})$. Its lower (upper) error bar is the maximum of the 16th (84th) percentile and $m=2$ (see text for details). We see that $m_{\rm fil}$ remains nearly constant at $R = 1~\rm kpc$. At $R = 2~\rm kpc$, it first decreases at $\beta = 10$ and then increases for $\beta \leq 1$.
    The dashed and dotted curves show $m_{\rm fil}$ estimated from the fastest growing mode of magneto-Jeans and magneto-Toomre instabilities, respectively, with the shaded envelopes representing their respective uncertainty range. Except for $\beta \ll 1$, and the $\beta = 10$ point at $R = 2~\rm kpc$, the simulations show agreement with predictions from the magneto-Jeans instability. }
    \label{fig:mFil_with_pbeta}
\end{figure}

We can compare the filament spacing of the galaxies with expectations from linear theory. For this purpose we use the peak of $A(m)$ (see \autoref{fig:mFil_with_pbeta}) to extract the number of filaments $m_{\rm fil}$, which is related to the filament spacing via $\lambda_{\rm fil}= 2\pi R/m_{\rm fil}$. We show how $m_{\rm fil}$ varies with initial $\beta$ in \autoref{fig:mFil_with_pbeta}. The circular points in the figure represent $m_{\rm fil}$ extracted from the simulations. The different colours correspond to the two $\rm 100~pc$-wide galactocentric bins centred at $R= { 1,2}~\rm kpc$, similar to \autoref{fig:a_m_with_time}. To account for the time evolution of $A(m)$ and thus of $m_\mathrm{fil}$, we calculate $m_{\rm fil}$ for each time snapshot for $t \in (t_{\mathrm{onset}}, t_{\mathrm{end}})$, and compute the median and 16th to 84th percentile range of the resulting values. We plot the medians as circular points, and the 16th to 84th percentile range as error bars, with a minimum error bar size corresponding to $m=2$, which is half the standard deviation of the Gaussian filter we use to smooth the $A(m)$ curves. We do not show the $\beta = \infty, \beta= 100$ runs at $R = 1~\rm kpc$ in this figure because they do not form filaments at that radius. We see that the variation of $m_{\rm fil}$ with $\beta$ clearly depends upon $R$. At $R = 1~\rm kpc$, $m_{\rm fil}$ is nearly independent of $\beta$, while at $R = 2~\rm kpc$, $m_{\rm fil}$ first decreases for intermediate magnetisation down to $\beta = 10$, rises for equipartition fields at $\beta = 1$, and then remains constant for $\beta \leq 1$.

For comparison we also show two curves in \autoref{fig:mFil_with_pbeta} predicted by linear theory. The dashed line represents the prediction of magneto-Jeans instability, where azimuthal magnetic fields work with self-gravity to destabilise the galaxy \citep[][and references therein]{kim_amplification_2001}. The dotted lines show the Toomre instability \citep{toomre_1964}, where we have included the effects of magnetic fields as an additional stabilising pressure term. We refer to this as the magneto-Toomre length scale hereafter. 

To generate these predictions, we start from the dispersion relations. For magneto-Jeans instability this is \citep[equation 23 in][]{kim_ostriker_2001}
\begin{equation}\label{eqn:dispersionMJI_dimensionfull}
    \begin{split}
    \gamma^{4} + \underbrace { \left [ \kappa^{2} - 2\pi G \Sigma_0|k| + (\cs^{2} + \va ^{2})k^{2} \right ] }_{\text{magneto-Toomre term}}  \gamma^2 + \\ \left [ \cs^{2} k^{2}(t) - 2\pi G\Sigma_{0}|k| \right ]\va^{2} k_y^{2} = 0,
    \end{split}
\end{equation}
where $\gamma$ is the growth rate of the instability and $k= \sqrt{k_x ^{2} + k_y^{2}}$ is the wavenumber. Here, $x$ is the local radial co-ordinate and $y$ is the local azimuthal co-ordinate\footnote{In principle, the radial wavenumber is time-dependent, with $k_x(t) = k_x + q\Omega k_y t$. This time dependence arises from the shearing-wavelets ansatz used in the derivation of the dispersion relation. Shearing-wavelets are travelling waves in the local frame of reference that shear ($q>0$) with the background flow of the galaxy. We neglect this time dependence since it only changes the filament number by $\leq 10\%$.} \citep{goldreich_lynden_bell_swing_1965}. The last term on the left-hand side is the magnetic destabilisation term, which is present only for non-zero magnetisation ($\va \neq 0$) and non-axisymmetric wave modes $k_y\neq 0$, and is responsible for growth ($\gamma>0$) of wave-modes that satisfy the Jeans criterion, $k\leq 2\pi G \Sigma_0/\cs^{2}$ \citep[see][for details]{kim_amplification_2001}. The dispersion relation for the magneto-Toomre instability is given by 
\begin{equation}\label{eqn:dispersionmagnetoToomre_dimensionfull}
    \gamma ^{2}  + \kappa^{2} - 2\pi G \Sigma_0|k| + (\cs^{2} + \va ^{2})k^{2} = 0, 
\end{equation}
which is the Toomre dispersion relation with an additional magnetic pressure term added to the thermal pressure. We see that this is the same as the co-efficient of the quadratic term in \autoref{eqn:dispersionMJI_dimensionfull}, which reduces to \autoref{eqn:dispersionmagnetoToomre_dimensionfull} if one only considers axisymmetric wave-modes with $k_y = 0$.

To get the filament number of the two instabilities at a particular galactocentric radius, we simply divide the circumference at that radius by the wavelength of the fastest growing mode of the respective instability. That is $m_{\mathrm{j}} = 2\pi R/\lambda_{j}$, where $j\in $ \{magneto-Jeans, magneto-Toomre\}. For the magneto-Jeans instability, we cannot solve for $\lambda_j$ exactly, but we can approximate it by $\lambda_{\mathrm{magneto-Jeans}} \approx  2\cs^{2}/G\Sigma$. We show in \aref{appendix:lengthScale_magneto-Jeans} that instead using a numerical solution for the location of the maximum does not lead to qualitatively different conclusions, so here we simply adopt the approximate analytic expression. For the magneto-Toomre instability, we can find the most unstable wavelength exactly: $\lambda_{\rm magneto-Toomre} = 2c_{\mathrm{eff}}^{2}/G\Sigma$, where $\cs$ is replaced by $c_{\rm eff}$ (c.f. \autoref{subsect:parameterStudy}) to include the stabilising effect of B-fields. In both the cases $\Sigma$ is the surface density of the galaxy. We estimate $\Sigma$ by using model galaxies that share their physical parameters with the initial conditions of our simulations. That is, galaxies with the same $Q_{\rm eff}$ profile (given by \autoref{fig:toomre_radialProfile}), but varying $\beta$. This gives $\Sigma = \kappa c_{\mathrm{eff}}/\pi G Q_{\rm eff}$. Substituting $\Sigma$ into the expressions for $m_{j}$ of the two dispersion relations gives 
\begin{align}\label{eqn:mToomre}
    m_{\mathrm{magneto-Jeans}} (\beta, R) &= R\frac{\kappa (R)}{Q_{\mathrm{eff}}(R) \xi}  \frac{ \sqrt{1 + 1/\beta}}{\cs }, \\ 
    m_{\mathrm{magneto-Toomre}} (\beta, R) &= R\frac{\kappa (R)}{Q_{\mathrm{eff}}(R) \xi}  \frac{1}{\cs \sqrt{1 + 1/\beta} }. 
\end{align}
Here, we have used $\qeff\xi$ instead of $\qeff$ in order to account for the non-zero disc thickness of the simulations. The dimensionless quantity $\xi$ is an empirical correction factor given by $\xi = 0.8 + 0.7(\sigma_z/\sigma_R)$ , first introduced in \cite{romeo_falstad_2013} (their equation~18). Here $\sigma_R, \sigma_z$ are the radial and vertical gas velocity dispersions, respectively. To estimate the velocity dispersions, we take the standard deviation of the radial and vertical velocities in the same radial bins that we use to estimate $m_{\rm fil}$. Since the velocity dispersions are  3D quantities, we extend the radial bins by $50~\rm pc$ in the direction perpendicular to the plane of the galaxy. To exclude the influence of the initial turbulent velocity fluctuations (see \autoref{subsect:simulationSetup}), we compute or velocity dispersions as time averages over the interval $(0,t_{\rm onset})$, i.e.~before the onset of filament formation. Since we find that the ratio $\sigma_z/\sigma_R$ varies by $\approx 10\%$ for all $\beta$ and $R$, we use its mean value of $\sigma_z/\sigma_R = 0.75$ in our analysis. This gives $\xi = 1.325$ for all our galaxies, and at both radii. Finally, we calculate the spread in $m$ by propagating the errors in $Q_{\rm eff}$, $\kappa(R)$, and $R$ due to the thickness of the radial bins.  

We see that the two $m_{\rm fil}$ analytical curves diverge with decreasing $\beta$; $m_{\rm magneto-Toomre}$ decreases while $m_{\rm magneto-Jeans}$ increases. Physically, the decrease in $m_{\rm magneto-Toomre}$ is due to the increase in magnetic pressure with decreasing $\beta$. The additional pressure, which acts to stabilise on smaller length scales, pushes the length scale of the instability to larger values (c.f.~\autoref{eqn:dispersionmagnetoToomre_dimensionfull}). Since this is balanced by magnetic tension in the magneto-Jeans instability, $m_{\rm magneto-Jeans}$ increases with decreasing (increasing) $\beta$ (surface density). At $R = 1~\rm kpc$, $m_{\rm fil} \sim m_{\rm magneto-Jeans}$ till $\beta = 1$, and then starts to disagree for $\beta \leq 1$. For $\beta\leq 1$, $m_\mathrm{magneto-Toomre}<m_{\rm fil}<m_{\mathrm{magneto-Jeans}}$. At $R = 2~\rm kpc$, we see that $m_{\rm fil}$ for low magnetisation $\beta \geq 100$ agrees with linear theory. This agreement for the $\beta = \infty$ case was also seen in \citetalias{arora_2025}. It is expected that dynamically insignificant magnetic fields with $\beta = 100$ do not change this behaviour. At $\beta = 10$, however, we see that $m_{\rm fil}$ dips below the two curves, being closer to $m_{\rm magneto-Toomre}$ than $m_{\rm magneto-Jeans}$. This changes for $\beta = 1$, where the simulations once again agree with $m_{\rm magneto-Jeans}$. For $\beta \leq 1$, we see $m_{\mathrm{magneto-Toomre}}<m_{\rm fil}<m_{\mathrm{magneto-Jeans}}$, similar to its behaviour at $R = 1~\rm kpc$. Thus, the variation of $m_{\rm fil}$ with $\beta$ displays a greater complexity than what is predicted by linear analysis. 

We can use \autoref{fig:growthRate_with_pbeta} to aid in the interpretation of the disagreements between theory and simulations. At $R = 2~\rm kpc$, the dip in $m_{\rm fil}$ for the $\beta = 10$ case that takes the simulation point closer to $m_{\rm magneto-Toomre}$ might be a consequence of magnetic fields acting to stabilise. We can see this in \autoref{fig:growthRate_with_pbeta}, where the filament growth rate at $\beta = 10$ and $R = 2.0~\rm kpc$ is equal to the growth rate at higher $\beta \in (100, \infty)$. Since these are galaxies that share their initial $Q_{\rm eff}$ parameter, magnetic fields tend to behave as an additional pressure term. Similar behaviour can be attributed to $\beta < 1$ at $R = 2~\rm kpc$, where the filament growth rate once again saturates (see \autoref{fig:growthRate_with_pbeta}). Hereafter, any additional magnetisation is associated with an increase in isotropic pressure. This interpretation, where a drop in $m_{\rm fil}$ is due to stabilising effects, however, is not applicable to how magnetic fields behave at $R = 1~\rm kpc$. At this radius, while the $m_{\rm fil}$ does not change with $\beta$, the filament growth rate shows a consistent increase. We suspect that non-linear effects might start dominating for this regime at $R \leq 1~\rm kpc$ and with $\beta < 1$. To investigate this further, one needs to systematically relax the physical assumptions made in deriving the magneto-Jeans dispersion relation. This will be part of a future study, where we focus on the linear stages of the instability (Arora et. al. in prep).

\section{Discussion} \label{sec:Discussion}
In this section we discuss our results in relation to existing literature and note the limitations of our approach. In \autoref{subsect:natureOfInstability} we compare our findings with linear instability analysis and numerical simulations. In \autoref{subsect: limitation_futureWork}, we outline the limitations of our simulations and suggest possible directions of future work. 

\subsection{Comparison with other works} \label{subsect:natureOfInstability}

\subsubsection{Linear Theory}

Linear stability analyses of magnetised galactic discs have revealed that magnetic fields can destabilise the disc via the magneto-Jeans instability \citep{parker_1966, elmegreen_1987_magnetic, elmegreen1991, gammie_1996, fan_lou_1997_mhdWavesSwing, kim_amplification_2001}. Moreover, it is expected that this is enhanced by the presence of the Parker instability, which acts perpendicular to the plane of the disc \citep{parker_1966, mouschovias_2009, elmegreen1991,tharakkal_2023_steady_states_PI}. A key aspect of the magneto-Jeans instability is that it depends critically upon the shear of the galactic background. \cite{kim_ostriker_2001} found that for regions within a galaxy having non-zero shear ($q > 0$), in-plane azimuthal magnetic fields in-fact stabilise the disc till a critical magnetic field strength, switching to destabilisation only for higher magnetic fields. For $q = 0$, azimuthal magnetic fields were found to always destabilise. We note that this effect, while being linear, is not captured by the dispersion relation given by \autoref{eqn:dispersionMJI_dimensionfull} (see Figure 1 in \citealt{kim_amplification_2001}). This qualitatively agrees with what we find in our simulations. As we saw in \autoref{subsect:filamentFormation_timescales}, for $R = 1~\rm kpc$, where our galaxies have weak shear (small $q$), we see that magnetic fields destabilise and increase the filament formation rates, starting at negligible initial magnetisation: $\beta \in (10, 100)$. For $R = 2~\rm kpc$, where shear is stronger (large $q$), magnetic fields work as a pressure term and stabilise till $\beta = 10$, and then start to destabilise for $\beta$ lower than $\beta \in (1, 10)$. 

However, we also see additional effects not captured by linear analysis. For the filament growth rates  presented in \autoref{subsect:filamentFormation_timescales}, we see a saturation of the growth rates for $\beta \leq 1$ at $R \geq 2~\rm kpc$, but we do not observe the same behaviour for $R\leq 1.5~\rm kpc$. At these smaller radii, the growth rate continues to increase with decreasing $\beta$ up to the lowest $\beta = 0.1$ we sample in our simulation suite. Similarly, in our comparison of filament number (spacing) with linear theory in 
\autoref{subsect:filamentspacing}, we find that the number of filaments in our simulations largely agrees with predictions from the magneto-Jeans instability, but that there are disagreements at $R = 2~\rm kpc$ for $\beta$ values where magnetic fields act as stabilising pressure terms, and $R = 1~\rm kpc$ and $\beta \ll 1$, where we suspect that non-linear effects dominate.

\subsubsection{Simulations} \label{subsubsect:comparisonWithSimulations}

A few previous studies have also aimed to isolate the role of magnetic fields on dense structure formation in galaxies using global simulations similar to ours. Some of these find that introducing magnetic fields reduces the density contrast of their filamentary structures, even for runs with $\beta\leq 1$ \citep{dobbs_price_2008, khoperskov_global_2018}. In these studies, magnetic fields are regarded as stabilising agents. There are others that report a flip in this behaviour beyond a threshold magnetic field strength, where the density contrast starts to increase with increasing magnetisation  \citep{bastian_2019}. This turnover happens for runs with $\beta \leq 0.5$. This critical $\beta$ lies a factor of $2$ outside the range we find for $R = 2~\rm kpc$ (see \autoref{subsect:filamentFormation_timescales}). Thus, both sets of studies seem to be in tension with one another and with our results. A few key differences in the initial conditions of these studies and their analysis methods, however, relieves the tension. \citet{dobbs_price_2008} and \citet{khoperskov_global_2018} both initialise their magnetised galaxies keeping the $Q_{g} = \kappa \cs /\pi G \Sigma$ constant. \cite{bastian_2019} do the same by keeping the $Q_{\rm eff} = \kappa c_{\rm eff}/\pi G \Sigma$ constant, similar to what we do in our simulation suite. Thus, their hydrodynamical and magnetic cases, even with the same initial $\beta$, are intrinsically different. The models used in \cite{bastian_2019} have a higher initial surface density than the ones used in \citet{dobbs_price_2008} and \citet{khoperskov_global_2018} by a factor of $\sqrt{1 +1/\beta}$ because they account for the additional magnetic pressure support. This explains why \cite{bastian_2019} observe magnetic destabilisation, while \citet{dobbs_price_2008} and \citet{khoperskov_global_2018} do not. Moreover, all of these studies exclusively restrict their analysis to a region where the rotation curve is fully flat with $q=1$. Such regions, as we saw in our results and as expect from linear analysis \citep{kim_amplification_2001}, will have a higher threshold for magnetic destabilisation. This also explains the higher critical $\beta$ reported by \cite{bastian_2019}. For completeness, we compare two simulations with $\beta = \{\infty, 1\}$ that share their initial $Q_{\rm g}$ in \aref{appendix:runsWithSameQg}. We show that magnetic destabilisation in the $\beta = 1$ run still dominates for regions with $R\leq 2.0~\rm kpc$ ($q\leq 0.5$).

One can also compare our result-that magnetic field can be both stabilising and destabilising-with earlier isolated disc simulations that include additional effects such as gas cooling and heating, star formation, and feedback processes. Simulations of stratified local galactic patches find that magnetic fields significantly reduce their star formation rate (SFR) \citep{kim_ostriker_2015, iffrig_hennebelle_2017, girichidis_2018, kim_wong_tigress_2021}, but it is unclear to what extent these results are comparable to those of global simulations, since, among other caveats, the local shearing box approximation used in these studies discards the long-range interaction of gravitational modes that may be important to our findings. Global simulations find that magnetic support has much more modest effects, with equipartition magnetic fields reducing the SFR by only a factor of $\sim 2$ \citep{steinwandel_2019,wibking_krumholz_2023,robinson_wadsley_2024}. Both groups of studies, however, agree and point towards magnetic stabilisation as the sole effect of magnetic fields. 

However, these simulations also focus their analysis on galactic regions where we find magnetic destabilisation to be sub-dominant. The local patches all set $q = 1$ \citep[e.g.][]{kim_wong_tigress_2021}, and the global studies focus on massive spiral galaxies where the fully-flat $q=1$ regions dominate due to their fast rising rotation curve. In the later case, studies risk averaging over the destabilising effects that are expected to be present in low-shear regions close to the galactic centre. For instance, we see that for global simulations done in \cite{robinson_wadsley_2024}, the central $2~\rm kpc$ region displays an increased star formation rate with increasing magnetisation (c.f.~their figure~5), the opposite of the trend they find at $R\geq 2~\rm kpc$, but because most star formation in this simulation occurs at larger galactocentric radii, the effect at larger radii dominates. Magnetic destabilisation is much more apparent in global dwarf galaxy simulations, which have a slow-rising rotation curve; \citep{whitworth_2025} show that the disc averaged SFR in their simulation remains unchanged in the presence of magnetic fields. On the other hand, it has been reported that including magnetic fields increases the dense, cold gas fractions in such simulations \citep{whitworth_smith_2023, gurman_steinwandel_2025, ryan_rowan_whitworth_2026}. We caution, however, that the SFR in such simulations might not trace the effects we find in our simulations. This is due to the inherent complexity of the sub-grid physics that couples with the instability. For instance, magnetically-enhanced winds in realistic simulations can decrease the SFR by reducing the fuel available for the process, opposing the destabilising effects seen in our simulations. 

\subsection{Caveats and future work} \label{subsect: limitation_futureWork}

Our simulations are controlled numerical experiments that explore the effects of magnetic fields on the global gravitational instability, and its role in the formation of filaments in disc galaxies. While our setup offers us advantages of isolating magnetic effects, and studying their dependence upon dimensionless parameters at a reasonable computational cost, we neglected some physical processes that are prevalent in galaxies and might directly influence these effects. First, we approximate the dark matter and stellar component via a static analytical potential. Doing so neglects their live gravitational interaction with the gas. The stellar mass field, for instance, has been shown to increase the gravitational destabilisation of gas \citep[e.g.,][]{jog_solomon_1984, jog_1992_swing, romeo_1992, romeo_wiegert_2011_q_stability}. It is unknown how the effects of the stellar potential couple with a magnetic fluid in 3D global simulations that start in equilibrium \citep[e.g.][]{thor_nexus2024}. The gravitational interaction of the disc with a dark matter halo can also incite bar formation in massive galaxies \citep[e.g.][]{sellwood1980_bars, sellwood2014_bar}, a process that is sped up in disc-dominated and turbulent, gas-rich galaxies \citep[][]{bland-Hawthorn_2023,bland-Hawthorn_2024}. As our results indicate, this interaction will potentially be influenced by magnetic fields. A detailed study of this effect is beyond the scope of the current work, and will be a part of a forthcoming study.

We also approximate the gas as isothermal, neglecting the chemistry and heating/cooling mechanisms that create a multiphase medium \citep{cox_2005_threePhase_ISM}. These processes might aid in the formation of over-densities within the filaments via the action of the thermal instability \citep{koyama_inutsuka_2000_thermal_instability_1D, vazquezsemadeni_molecular_2007}. Moreover, we also neglect the process of star formation, and various feedback mechanisms associated with it. These include supernovae, winds from massive stars, and the injection and presence of cosmic rays. A detailed analysis of their coupling remains out of the scope for this study. Future investigations will aim to incorporate them in order to build more realistic galaxy models.

\section{Conclusions} \label{sec:conclusions}

We perform a series of simulations of magnetised, self-gravitating, isothermal and isolated disc galaxies that are initialised in equilibrium. The study is aimed at exploring the effects of magnetic fields on the gas gravitational instability by quantifying their effects on filament/feather formation. We proceed by initialising a gravitationally unstable galaxy containing regions with $\qgas<1$ that are unstable to filament formation, which we simulate at varying initial ratio of thermal to magnetic pressure, $\beta \in \{ 0.1, 0.5, 1, 10, 100, \infty\}$, while keeping the effective Toomre parameter of the galaxy, $Q_{\rm eff}$ (c.f.~\autoref{eqn:q_eff}), constant. This ensures that we balance any additional magnetic pressure support by an analogous increase in the surface density and self-gravity of the galaxy. We analyse the dependence of filament morphology, growth rates and filament number on $\beta$ in the simulations. Our main conclusions are:

\begin{itemize}
    \item Our simulations show significant differences in filament growth rates, spacing, and morphology with $\beta$. This variation shows a systematic dependence upon the galactocentric radius $R$, or equivalently the galactic shear $q$. Thus, magnetic fields alter the gas gravitational instability and do not act as mere pressure terms. \\
    
    \item Instead, magnetic fields switch between stabilising pressure terms and destabilising agents. The filament growth rates rise approximately $\propto \log_{10}(1/\beta)$ for $\beta\leq 10$ at $R = 1.0, 1.5 ~\rm kpc$ where the rotation curve is close to solid-body ($q = 0.2, 0.36$), while at larger radii $R = 2, 2.5~\mathrm{kpc}$ where the rotaiton curve is closer to flat $(q = 0.5, 0.6)$ filament growth starts at lower $\beta$ and occurs for a narrower range, with $\beta \in \{1, 10 \}$. For $\beta$ outside this range, magnetic fields act as pressure terms and the growth rates show negligible variation. This switch between destabilisation and stabilisation, and the decrease in threshold $\beta$ for magnetic destabilisation with increasing $R$ (or $q$), qualitatively agrees with the predicted behaviour of magneto-Jeans instability \citep{kim_amplification_2001}. \\
    
    \item We find that the number of filaments along the azimuth, $m_{\rm fil}$, also varies with $\beta$ and $R$ (or $q$). At $R = 1.0 ~\rm kpc$, $m_{\rm fil}$ is independent of $\beta$. At $R = 2.0~\rm kpc$, $m_{\rm fil}$ shows a factor of $2$ variation across the $\beta$ values sampled in our work. Over most of this range the variation in $m_\mathrm{fil}$ (and thus in the inter-filament spacing) with local galactic conditions is consistent with the predictions for the fastest-growing unstable mode from linear analysis of the magneto-Jeans instability, with exceptions at $\beta \ll 1$ and in isolated other cases; in these cases we suspect that non-linear interactions modify the initial linear response in a way that changes the filament spacing.
    
\end{itemize}

Our simulations show that the gas gravitational instability in galaxies is modified by the inclusion of magnetic fields. Instead of acting as mere pressure terms, we find that magnetic fields can instead aid gravitational destabilisation. This effect starts being active above a threshold magnetic field strength, which increases with increasing background shear. This largely agrees with expectations from previous analytical work. While our simulations offered us the advantage of bridging linear theory with global galactic simulations that go into its non-linear stages, we are yet to understand the action of magnetic gravitational instability in real galaxies that have a stellar, dark matter component and undergo star formation. Future studies will aim to address this using the tools of linear analysis and controlled simulations with live stellar, dark matter components (Arora et al., in prep). These steps will allow us a more complete picture of the interplay between magnetic fields and gravity on galactic scales. 

\section*{Acknowledgements}

R.~A.~ heartfully thanks Alessandro Romeo, Joss Bland-Hawthorn and Shivan Khullar for insightful discussions, and acknowledges support from the LMK foundation. O.~A.~ acknowledges support from the Knut and Alice Wallenberg Foundation, the Swedish Research Council (grant 2025-04892), the Swedish National Space Agency (SNSA grants 2023-00164 and 2025-00405), the LMK foundation, and eSSENCE, a Swedish strategic research programme in e-Science. C.~F.~acknowledges funding provided by the Australian Research Council (Discovery Projects DP230102280 and DP250101526), and the Australia-Germany Joint Research Cooperation Scheme (UA-DAAD). M.~R.~K.~acknowledges support from the Australian Research Council through Laureate Fellowship FL220100020. We further acknowledge high-performance computing resources provided by the Leibniz Rechenzentrum and the Gauss Centre for Supercomputing (grants~pr32lo, pr48pi and GCS Large-scale project~10391), the Australian National Computational Infrastructure (grant~ek9) and the Pawsey Supercomputing Centre (project~pawsey0810) in the framework of the National Computational Merit Allocation Scheme and the ANU Merit Allocation Scheme. The simulation software, \texttt{FLASH}, was in part developed by the Flash Centre for Computational Science at the University of Chicago and the Department of Physics and Astronomy at the University of Rochester.

\section*{Data Availability}

The snapshots of the simulations will be shared upon reasonable requests from the authors.



\bibliographystyle{mnras}
\bibliography{arora_docto_library_Jun2026, magnetic_feather}

@article{ejdetjarn_2022,
	adsurl = {https://ui.adsabs.harvard.edu/abs/2022MNRAS.514..480E},
	archiveprefix = {arXiv},
	author = {{Ejdetj{\"a}rn}, Timmy and {Agertz}, Oscar and {{\"O}stlin}, G{\"o}ran and {Renaud}, Florent and {Romeo}, Alessandro B.},
	doi = {10.1093/mnras/stac1414},
	eprint = {2111.09322},
	journal = {\mnras},
	month = jul,
	number = {1},
	pages = {480-496},
	primaryclass = {astro-ph.GA},
	title = {{From giant clumps to clouds - III. The connection between star formation and turbulence in the ISM}},
	volume = {514},
	year = 2022}

@article{meidt_2024,
	adsurl = {https://ui.adsabs.harvard.edu/abs/2024ApJ...966...62M},
	archiveprefix = {arXiv},
	author = {{Meidt}, Sharon E. and {van der Wel}, Arjen},
	doi = {10.3847/1538-4357/ad12c5},
	eid = {62},
	eprint = {2312.02618},
	journal = {\apj},
	month = may,
	number = {1},
	pages = {62},
	primaryclass = {astro-ph.GA},
	title = {{Bottom's Dream and the Amplification of Filamentary Gas Structures and Stellar Spiral Arms}},
	volume = {966},
	year = 2024}

@article{fuchs_2001,
	adsurl = {https://ui.adsabs.harvard.edu/abs/2001A&A...368..107F},
	archiveprefix = {arXiv},
	author = {{Fuchs}, B.},
	doi = {10.1051/0004-6361:20000562},
	eprint = {astro-ph/0012458},
	journal = {\aap},
	month = mar,
	pages = {107-121},
	primaryclass = {astro-ph},
	title = {{Density waves in the shearing sheet. I. Swing amplification}},
	volume = {368},
	year = 2001}

@article{arora_2023,
	adsurl = {https://ui.adsabs.harvard.edu/abs/2024A&A...687A.276A},
	archiveprefix = {arXiv},
	author = {{Arora}, Raghav and {Federrath}, Christoph and {Banerjee}, Robi and {K{\"o}rtgen}, Bastian},
	doi = {10.1051/0004-6361/202348719},
	eid = {A276},
	eprint = {2311.16266},
	journal = {\aap},
	month = jul,
	pages = {A276},
	primaryclass = {astro-ph.GA},
	title = {{Role of magnetic fields in disc galaxies: spiral arm instability}},
	volume = {687},
	year = 2024}

@article{toomre_1964,
	adsurl = {https://ui.adsabs.harvard.edu/abs/1964ApJ...139.1217T},
	author = {{Toomre}, A.},
	doi = {10.1086/147861},
	journal = {\apj},
	month = may,
	pages = {1217-1238},
	title = {{On the gravitational stability of a disk of stars.}},
	volume = {139},
	year = 1964}

@book{binney_tremaine_1987_galactic_dynamics,
	adsurl = {https://ui.adsabs.harvard.edu/abs/1987gady.book.....B},
	author = {{Binney}, James and {Tremaine}, Scott},
	title = {{Galactic dynamics}},
	year = 1987}

@article{romeo_falstad_2013,
	adsurl = {https://ui.adsabs.harvard.edu/abs/2013MNRAS.433.1389R},
	archiveprefix = {arXiv},
	author = {{Romeo}, Alessandro B. and {Falstad}, Niklas},
	doi = {10.1093/mnras/stt809},
	eprint = {1302.4291},
	journal = {\mnras},
	month = aug,
	number = {2},
	pages = {1389-1397},
	primaryclass = {astro-ph.CO},
	title = {{A simple and accurate approximation for the Q stability parameter in multicomponent and realistically thick discs}},
	volume = {433},
	year = 2013}

@article{griv_wang_2014_hydro_sims_twoD,
	adsurl = {https://ui.adsabs.harvard.edu/abs/2014NewA...30....8G},
	author = {{Griv}, Evgeny and {Wang}, Hsiang-Hsu},
	doi = {10.1016/j.newast.2014.01.001},
	journal = {\na},
	month = jul,
	pages = {8-27},
	title = {{Density wave formation in differentially rotating disk galaxies: Hydrodynamic simulation of the linear regime}},
	volume = {30},
	year = 2014}

@article{kim_ostriker_2002,
	adsurl = {https://ui.adsabs.harvard.edu/abs/2002ApJ...570..132K},
	archiveprefix = {arXiv},
	author = {{Kim}, Woong-Tae and {Ostriker}, Eve C.},
	doi = {10.1086/339352},
	eprint = {astro-ph/0111398},
	journal = {\apj},
	month = may,
	number = {1},
	pages = {132-151},
	primaryclass = {astro-ph},
	title = {{Formation and Fragmentation of Gaseous Spurs in Spiral Galaxies}},
	volume = {570},
	year = 2002}

@article{kim_wong_kim_2015,
	adsurl = {https://ui.adsabs.harvard.edu/abs/2015ApJ...809...33K},
	archiveprefix = {arXiv},
	author = {{Kim}, Yonghwi and {Kim}, Woong-Tae and {Elmegreen}, Bruce G.},
	doi = {10.1088/0004-637X/809/1/33},
	eid = {33},
	eprint = {1506.07178},
	journal = {\apj},
	month = aug,
	number = {1},
	pages = {33},
	primaryclass = {astro-ph.GA},
	title = {{Wiggle Instability of Galactic Spiral Shocks: Effects of Magnetic Fields}},
	volume = {809},
	year = 2015}

@article{kim_wong_tigress_2021,
	adsurl = {https://ui.adsabs.harvard.edu/abs/2020ApJ...898...35K},
	archiveprefix = {arXiv},
	author = {{Kim}, Woong-Tae and {Kim}, Chang-Goo and {Ostriker}, Eve C.},
	doi = {10.3847/1538-4357/ab9b87},
	eid = {35},
	eprint = {2006.05614},
	journal = {\apj},
	month = jul,
	number = {1},
	pages = {35},
	primaryclass = {astro-ph.GA},
	title = {{Local Simulations of Spiral Galaxies with the TIGRESS Framework. I. Star Formation and Arm Spurs/Feathers}},
	volume = {898},
	year = 2020}

@article{bastian_2019,
	adsurl = {https://ui.adsabs.harvard.edu/abs/2019MNRAS.489.5004K},
	archiveprefix = {arXiv},
	author = {{K{\"o}rtgen}, Bastian and {Banerjee}, Robi and {Pudritz}, Ralph E. and {Schmidt}, Wolfram},
	doi = {10.1093/mnras/stz2491},
	eprint = {1909.01623},
	journal = {\mnras},
	month = nov,
	number = {4},
	pages = {5004-5021},
	primaryclass = {astro-ph.GA},
	title = {{Global dynamics of the interstellar medium in magnetized disc galaxies}},
	volume = {489},
	year = 2019}

@article{lang_meidt_2020,
	adsurl = {https://ui.adsabs.harvard.edu/abs/2020ApJ...897..122L},
	archiveprefix = {arXiv},
	author = {{Lang}, Philipp and {Meidt}, Sharon E. and {Rosolowsky}, Erik and {Nofech}, Joseph and {Schinnerer}, Eva and {Leroy}, Adam K. and {Emsellem}, Eric and {Pessa}, Ismael and {Glover}, Simon C.~O. and {Groves}, Brent and {Hughes}, Annie and {Kruijssen}, J.~M. Diederik and {Querejeta}, Miguel and {Schruba}, Andreas and {Bigiel}, Frank and {Blanc}, Guillermo A. and {Chevance}, M{\'e}lanie and {Colombo}, Dario and {Faesi}, Christopher and {Henshaw}, Jonathan D. and {Herrera}, Cinthya N. and {Liu}, Daizhong and {Pety}, J{\'e}r{\^o}me and {Puschnig}, Johannes and {Saito}, Toshiki and {Sun}, Jiayi and {Usero}, Antonio},
	doi = {10.3847/1538-4357/ab9953},
	eid = {122},
	eprint = {2005.11709},
	journal = {\apj},
	month = jul,
	number = {2},
	pages = {122},
	primaryclass = {astro-ph.GA},
	title = {{PHANGS CO Kinematics: Disk Orientations and Rotation Curves at 150 pc Resolution}},
	volume = {897},
	year = 2020}

@article{beck_chamandy_elson_2020,
	article-number = {4},
	author = {Beck, Rainer and Chamandy, Luke and Elson, Ed and Blackman, Eric G.},
	doi = {10.3390/galaxies8010004},
	issn = {2075-4434},
	journal = {Galaxies},
	number = {1},
	title = {Synthesizing Observations and Theory to Understand Galactic Magnetic Fields: Progress and Challenges},
	url = {https://www.mdpi.com/2075-4434/8/1/4},
	volume = {8},
	year = {2020}}

@article{beck_2015,
	adsurl = {https://ui.adsabs.harvard.edu/abs/2015A&ARv..24....4B},
	archiveprefix = {arXiv},
	author = {{Beck}, Rainer},
	doi = {10.1007/s00159-015-0084-4},
	eid = {4},
	eprint = {1509.04522},
	journal = {\aapr},
	month = dec,
	pages = {4},
	primaryclass = {astro-ph.GA},
	title = {{Magnetic fields in spiral galaxies}},
	volume = {24},
	year = 2015}

@article{dobbs_price_2008,
	adsurl = {https://ui.adsabs.harvard.edu/abs/2008MNRAS.383..497D},
	archiveprefix = {arXiv},
	author = {{Dobbs}, C.~L. and {Price}, D.~J.},
	doi = {10.1111/j.1365-2966.2007.12591.x},
	eprint = {0710.3558},
	journal = {\mnras},
	month = jan,
	number = {2},
	pages = {497-512},
	primaryclass = {astro-ph},
	title = {{Magnetic fields and the dynamics of spiral galaxies}},
	volume = {383},
	year = 2008}

@article{inoue_2018,
	adsurl = {https://ui.adsabs.harvard.edu/abs/2018MNRAS.474.3466I},
	archiveprefix = {arXiv},
	author = {{Inoue}, Shigeki and {Yoshida}, Naoki},
	doi = {10.1093/mnras/stx2978},
	eprint = {1706.01895},
	journal = {\mnras},
	month = mar,
	number = {3},
	pages = {3466-3487},
	primaryclass = {astro-ph.GA},
	title = {{Spiral-arm instability: giant clump formation via fragmentation of a galactic spiral arm}},
	volume = {474},
	year = 2018}

@article{inoue_2019,
	adsurl = {https://ui.adsabs.harvard.edu/abs/2019MNRAS.485.3024I},
	archiveprefix = {arXiv},
	author = {{Inoue}, Shigeki and {Yoshida}, Naoki},
	doi = {10.1093/mnras/stz584},
	eprint = {1807.02988},
	journal = {\mnras},
	month = may,
	number = {3},
	pages = {3024-3041},
	primaryclass = {astro-ph.GA},
	title = {{Spiral-arm instability - II. Magnetic destabilization}},
	volume = {485},
	year = 2019}

@book{shukurov_subramanian_2021,
	author = {Shukurov, Anvar and Subramanian, Kandaswamy},
	collection = {Cambridge Astrophysics},
	doi = {10.1017/9781139046657},
	place = {Cambridge},
	publisher = {Cambridge University Press},
	series = {Cambridge Astrophysics},
	title = {Astrophysical Magnetic Fields: From Galaxies to the Early Universe},
	year = {2021}}

@article{parker_1966,
	adsurl = {https://ui.adsabs.harvard.edu/abs/1966ApJ...145..811P},
	author = {{Parker}, E.~N.},
	doi = {10.1086/148828},
	journal = {\apj},
	month = sep,
	pages = {811},
	title = {{The Dynamical State of the Interstellar Gas and Field}},
	volume = {145},
	year = 1966}

@article{mouschovias_2009,
	adsurl = {https://ui.adsabs.harvard.edu/abs/2009MNRAS.397...14M},
	archiveprefix = {arXiv},
	author = {{Mouschovias}, Telemachos Ch. and {Kunz}, Matthew W. and {Christie}, Duncan A.},
	doi = {10.1111/j.1365-2966.2009.14472.x},
	eprint = {0901.0914},
	journal = {\mnras},
	month = jul,
	number = {1},
	pages = {14-23},
	primaryclass = {astro-ph.GA},
	title = {{Formation of interstellar clouds: Parker instability with phase transitions}},
	volume = {397},
	year = 2009}

@inproceedings{dubey_fisher_2008,
	adsurl = {https://ui.adsabs.harvard.edu/abs/2008ASPC..385..145D},
	author = {{Dubey}, A. and {Fisher}, R. and {Graziani}, C. and {Jordan}, G.~C., IV and {Lamb}, D.~Q. and {Reid}, L.~B. and {Rich}, P. and {Sheeler}, D. and {Townsley}, D. and {Weide}, K.},
	booktitle = {Numerical Modeling of Space Plasma Flows},
	editor = {{Pogorelov}, N.~V. and {Audit}, E. and {Zank}, G.~P.},
	month = apr,
	pages = {145},
	series = {Astronomical Society of the Pacific Conference Series},
	title = {{Challenges of Extreme Computing using the FLASH code}},
	volume = {385},
	year = 2008}

@article{Safronov_1960,
	adsurl = {https://ui.adsabs.harvard.edu/abs/1960AnAp...23..979S},
	author = {{Safronov}, V.~S.},
	journal = {Annales d'Astrophysique},
	month = feb,
	pages = {979},
	title = {{On the gravitational instability in flattened systems with axial symmetry and non-uniform rotation}},
	volume = {23},
	year = 1960}

@article{jog_solomon_1984,
	adsurl = {https://ui.adsabs.harvard.edu/abs/1984ApJ...276..114J},
	author = {{Jog}, C.~J. and {Solomon}, P.~M.},
	doi = {10.1086/161597},
	journal = {\apj},
	month = jan,
	pages = {114-126},
	title = {{Two-fluid gravitational instabilities in a galactic disk}},
	volume = {276},
	year = 1984}

@article{goldreich_lynden_bell_swing_1965,
	adsurl = {https://ui.adsabs.harvard.edu/abs/1965MNRAS.130..125G},
	author = {{Goldreich}, P. and {Lynden-Bell}, D.},
	doi = {10.1093/mnras/130.2.125},
	journal = {\mnras},
	month = jan,
	pages = {125},
	title = {{II. Spiral arms as sheared gravitational instabilities}},
	volume = {130},
	year = 1965}

@article{julian_toomre_1966,
	adsurl = {https://ui.adsabs.harvard.edu/abs/1966ApJ...146..810J},
	author = {{Julian}, William H. and {Toomre}, Alar},
	doi = {10.1086/148957},
	journal = {\apj},
	month = dec,
	pages = {810},
	title = {{Non-Axisymmetric Responses of Differentially Rotating Disks of Stars}},
	volume = {146},
	year = 1966}

@article{binney_2020,
	adsurl = {https://ui.adsabs.harvard.edu/abs/2020MNRAS.496..767B},
	archiveprefix = {arXiv},
	author = {{Binney}, James},
	doi = {10.1093/mnras/staa1485},
	eprint = {1906.11696},
	journal = {\mnras},
	month = jul,
	number = {1},
	pages = {767-783},
	primaryclass = {astro-ph.GA},
	title = {{The shearing sheet and swing amplification revisited}},
	volume = {496},
	year = 2020}

@article{lin_shu_1964,
	adsurl = {https://ui.adsabs.harvard.edu/abs/1964ApJ...140..646L},
	author = {{Lin}, C.~C. and {Shu}, Frank H.},
	doi = {10.1086/147955},
	journal = {\apj},
	month = aug,
	pages = {646},
	title = {{On the Spiral Structure of Disk Galaxies.}},
	volume = {140},
	year = 1964}

@article{brucy_2023,
	adsurl = {https://ui.adsabs.harvard.edu/abs/2023A&A...675A.144B},
	archiveprefix = {arXiv},
	author = {{Brucy}, No{\'e} and {Hennebelle}, Patrick and {Colman}, Tine and {Iteanu}, Simon},
	doi = {10.1051/0004-6361/202244915},
	eid = {A144},
	eprint = {2305.18012},
	journal = {\aap},
	month = jul,
	pages = {A144},
	primaryclass = {astro-ph.GA},
	title = {{Large-scale turbulent driving regulates star formation in high-redshift gas-rich galaxies. II. Influence of the magnetic field and the turbulent compressive fraction}},
	volume = {675},
	year = 2023}

@article{steinwandel_2019,
	adsurl = {https://ui.adsabs.harvard.edu/abs/2019MNRAS.483.1008S},
	archiveprefix = {arXiv},
	author = {{Steinwandel}, U.~P. and {Beck}, M.~C. and {Arth}, A. and {Dolag}, K. and {Moster}, B.~P. and {Nielaba}, P.},
	doi = {10.1093/mnras/sty3083},
	eprint = {1808.09975},
	journal = {\mnras},
	month = feb,
	number = {1},
	pages = {1008-1028},
	primaryclass = {astro-ph.GA},
	title = {{Magnetic buoyancy in simulated galactic discs with a realistic circumgalactic medium}},
	volume = {483},
	year = 2019}

@article{cox_2005_threePhase_ISM,
	adsurl = {https://ui.adsabs.harvard.edu/abs/2005ARA&A..43..337C},
	author = {{Cox}, Donald P.},
	doi = {10.1146/annurev.astro.43.072103.150615},
	journal = {\araa},
	month = sep,
	number = {1},
	pages = {337-385},
	title = {{The Three-Phase Interstellar Medium Revisited}},
	volume = {43},
	year = 2005}

@book{choudhuri_rai_1998,
	adsurl = {https://ui.adsabs.harvard.edu/abs/1998pfp..book.....C},
	author = {{Choudhuri}, Arnab Rai},
	doi = {10.1017/CBO9781139171069},
	publisher = {"Cambridge University Press"},
	title = {{The Physics of Fluids and Plasmas: An Introduction for Astrophysicists}},
	year = 1998}

@article{balbus_hawley_1998_MRI_review,
	adsurl = {https://ui.adsabs.harvard.edu/abs/1998RvMP...70....1B},
	author = {{Balbus}, Steven A. and {Hawley}, John F.},
	doi = {10.1103/RevModPhys.70.1},
	journal = {Reviews of Modern Physics},
	month = jan,
	number = {1},
	pages = {1-53},
	title = {{Instability, turbulence, and enhanced transport in accretion disks}},
	volume = {70},
	year = 1998}

@article{kolmogorov_1941,
	adsurl = {https://ui.adsabs.harvard.edu/abs/1941DoSSR..30..301K},
	author = {{Kolmogorov}, A.},
	journal = {Akademiia Nauk SSSR Doklady},
	month = jan,
	pages = {301-305},
	title = {{The Local Structure of Turbulence in Incompressible Viscous Fluid for Very Large Reynolds' Numbers}},
	volume = {30},
	year = 1941}

@article{tharakkal_2023_steady_states_PI,
	adsurl = {https://ui.adsabs.harvard.edu/abs/2023MNRAS.525.2972T},
	archiveprefix = {arXiv},
	author = {{Tharakkal}, Devika and {Shukurov}, Anvar and {Gent}, Frederick A. and {Sarson}, Graeme R. and {Snodin}, Andrew},
	doi = {10.1093/mnras/stad2475},
	eprint = {2305.03318},
	journal = {\mnras},
	month = oct,
	number = {2},
	pages = {2972-2984},
	primaryclass = {astro-ph.GA},
	title = {{Steady states of the Parker instability: the effects of rotation}},
	volume = {525},
	year = 2023}

@article{koyama_inutsuka_2000_thermal_instability_1D,
	adsurl = {https://ui.adsabs.harvard.edu/abs/2000ApJ...532..980K},
	archiveprefix = {arXiv},
	author = {{Koyama}, Hiroshi and {Inutsuka}, Shu-Ichiro},
	doi = {10.1086/308594},
	eprint = {astro-ph/9912509},
	journal = {\apj},
	month = apr,
	number = {2},
	pages = {980-993},
	primaryclass = {astro-ph},
	title = {{Molecular Cloud Formation in Shock-compressed Layers}},
	volume = {532},
	year = 2000}

@article{TrueloveEtAl1997,
	adsurl = {http://adsabs.harvard.edu/abs/1997ApJ...489L.179T},
	author = {{Truelove}, J.~K. and {Klein}, R.~I. and {McKee}, C.~F. and {Holliman}, II, J.~H. and {Howell}, L.~H. and {Greenough}, J.~A.},
	doi = {10.1086/316779},
	journal = {\apjl},
	month = nov,
	pages = {L179},
	title = {{The Jeans Condition: A New Constraint on Spatial Resolution in Simulations of Isothermal Self-gravitational Hydrodynamics}},
	volume = 489,
	year = 1997}

@article{FryxellEtAl2000,
	adsurl = {http://cdsads.u-strasbg.fr/abs/2000ApJS..131..273F},
	author = {{Fryxell}, B. and {Olson}, K. and {Ricker}, P. and {Timmes}, F.~X. and {Zingale}, M. and {Lamb}, D.~Q. and {MacNeice}, P. and {Rosner}, R. and {Truran}, J.~W. and {Tufo}, H.},
	doi = {10.1086/317361},
	journal = {\apjs},
	month = nov,
	pages = {273-334},
	title = {{FLASH: An Adaptive Mesh Hydrodynamics Code for Modeling Astrophysical Thermonuclear Flashes}},
	volume = 131,
	year = 2000}

@article{KrumholzMcKee2005,
	adsurl = {http://cdsads.u-strasbg.fr/abs/2005ApJ...630..250K},
	author = {{Krumholz}, M.~R. and {McKee}, C.~F.},
	doi = {10.1086/431734},
	eprint = {arXiv:astro-ph/0505177},
	journal = {\apj},
	month = sep,
	pages = {250-268},
	title = {{A General Theory of Turbulence-regulated Star Formation, from Spirals to Ultraluminous Infrared Galaxies}},
	volume = 630,
	year = 2005}

@article{TaskerTan2009,
	adsurl = {http://adsabs.harvard.edu/abs/2009ApJ...700..358T},
	archiveprefix = {arXiv},
	author = {{Tasker}, E.~J. and {Tan}, J.~C.},
	doi = {10.1088/0004-637X/700/1/358},
	eprint = {0811.0207},
	journal = {\apj},
	month = jul,
	pages = {358-375},
	title = {{Star Formation in Disk Galaxies. I. Formation and Evolution of Giant Molecular Clouds via Gravitational Instability and Cloud Collisions}},
	volume = 700,
	year = 2009}

@article{BournaudEtAl2010,
	adsurl = {http://adsabs.harvard.edu/abs/2010MNRAS.409.1088B},
	archiveprefix = {arXiv},
	author = {{Bournaud}, F. and {Elmegreen}, B.~G. and {Teyssier}, R. and {Block}, D.~L. and {Puerari}, I.},
	doi = {10.1111/j.1365-2966.2010.17370.x},
	eprint = {1007.2566},
	journal = {\mnras},
	month = dec,
	pages = {1088-1099},
	primaryclass = {astro-ph.CO},
	title = {{ISM properties in hydrodynamic galaxy simulations: turbulence cascades, cloud formation, role of gravity and feedback}},
	volume = 409,
	year = 2010}

@article{PadoanNordlund2011,
	archiveprefix = {arXiv},
	author = {{Padoan}, P. and {Nordlund}, {\AA}.},
	doi = {10.1088/0004-637X/730/1/40},
	eprint = {0907.0248},
	journal = {\apj},
	month = mar,
	pages = {40},
	primaryclass = {astro-ph.GA},
	title = {{The Star Formation Rate of Supersonic Magnetohydrodynamic Turbulence}},
	volume = 730,
	year = 2011}

@article{WaaganFederrathKlingenberg2011,
	adsurl = {http://adsabs.harvard.edu/abs/2011JCoPh.230.3331W},
	author = {{Waagan}, K. and {Federrath}, C. and {Klingenberg}, C.},
	doi = {10.1016/j.jcp.2011.01.026},
	journal = JournalofComputationalPhysics,
	month = may,
	pages = {3331-3351},
	primaryclass = {astro-ph.IM},
	title = {{A robust numerical scheme for highly compressible magnetohydrodynamics: Nonlinear stability, implementation and tests}},
	volume = 230,
	year = 2011}

@article{FederrathEtAl2011,
	adsurl = {http://adsabs.harvard.edu/abs/2011PhRvL.107k4504F},
	archiveprefix = {arXiv},
	author = {{Federrath}, C. and {Chabrier}, G. and {Schober}, J. and {Banerjee}, R. and {Klessen}, R.~S. and {Schleicher}, D.~R.~G.},
	doi = {10.1103/PhysRevLett.107.114504},
	eid = {114504},
	eprint = {1109.1760},
	journal = PhysicalReviewLetters,
	month = sep,
	number = 11,
	pages = {114504},
	primaryclass = {physics.flu-dyn},
	title = {{Mach Number Dependence of Turbulent Magnetic Field Amplification: Solenoidal versus Compressive Flows}},
	volume = 107,
	year = 2011}

@article{HennebelleChabrier2011,
	adsurl = {http://adsabs.harvard.edu/abs/2011ApJ...743L..29H},
	archiveprefix = {arXiv},
	author = {{Hennebelle}, P. and {Chabrier}, G.},
	doi = {10.1088/2041-8205/743/2/L29},
	eid = {L29},
	eprint = {1110.0033},
	journal = {\apjl},
	month = dec,
	pages = {L29},
	primaryclass = {astro-ph.GA},
	title = {{Analytical Star Formation Rate from Gravoturbulent Fragmentation}},
	volume = 743,
	year = 2011}

@article{FederrathKlessen2012,
	adsurl = {http://adsabs.harvard.edu/abs/2012ApJ...761..156F},
	archiveprefix = {arXiv},
	author = {{Federrath}, C. and {Klessen}, R.~S.},
	doi = {10.1088/0004-637X/761/2/156},
	eid = {156},
	eprint = {1209.2856},
	journal = {\apj},
	month = dec,
	pages = {156},
	primaryclass = {astro-ph.SR},
	title = {{The Star Formation Rate of Turbulent Magnetized Clouds: Comparing Theory, Simulations, and Observations}},
	volume = 761,
	year = 2012}

@article{Beck2016,
	adsurl = {http://adsabs.harvard.edu/abs/2016A%26ARv..24....4B},
	archiveprefix = {arXiv},
	author = {{Beck}, R.},
	doi = {10.1007/s00159-015-0084-4},
	eid = {4},
	eprint = {1509.04522},
	journal = {\aapr},
	month = dec,
	pages = {4},
	title = {{Magnetic fields in spiral galaxies}},
	volume = 24,
	year = 2016}

@article{SteinwandelEtAl2020,
	adsurl = {https://ui.adsabs.harvard.edu/abs/2020MNRAS.494.4393S},
	archiveprefix = {arXiv},
	author = {{Steinwandel}, Ulrich P. and {Dolag}, Klaus and {Lesch}, Harald and {Moster}, Benjamin P. and {Burkert}, Andreas and {Prieto}, Almudena},
	doi = {10.1093/mnras/staa817},
	eprint = {1907.11727},
	journal = {\mnras},
	month = may,
	number = {3},
	pages = {4393-4412},
	primaryclass = {astro-ph.GA},
	title = {{On the origin of magnetic driven winds and the structure of the galactic dynamo in isolated galaxies}},
	volume = {494},
	year = 2020}

@article{kim_amplification_2001,
	author = {Kim, Woong-Tae and Ostriker, Eve C.},
	doi = {10.1086/322330},
	issn = {0004-637X, 1538-4357},
	journal = {The Astrophysical Journal},
	language = {en},
	month = sep,
	note = {arXiv:astro-ph/0105375},
	number = {1},
	pages = {70--95},
	shorttitle = {Amplification, {Saturation}, and {Q} {Thresholds} for {Runaway}},
	title = {Amplification, {Saturation}, and {Q} {Thresholds} for {Runaway}: {Growth} of {Self}-{Gravitating} {Structures} in {Models} of {Magnetized} {Galactic} {Gas} {Disks}},
	url = {http://arxiv.org/abs/astro-ph/0105375},
	urldate = {2023-06-02},
	volume = {559},
	year = {2001}}

@article{wang_equilibrium_2010,
	adsurl = {https://ui.adsabs.harvard.edu/abs/2010MNRAS.407..705W},
	archiveprefix = {arXiv},
	author = {{Wang}, Hsiang-Hsu and {Klessen}, Ralf S. and {Dullemond}, Cornelis P. and {van den Bosch}, Frank C. and {Fuchs}, Burkhard},
	doi = {10.1111/j.1365-2966.2010.16942.x},
	eprint = {1004.5593},
	journal = {\mnras},
	month = sep,
	number = {2},
	pages = {705-720},
	primaryclass = {astro-ph.GA},
	title = {{Equilibrium initialization and stability of three-dimensional gas discs}},
	volume = {407},
	year = 2010}

@misc{borlaff_extragalactic_2023,
	author = {Borlaff, Alejandro S. and Lopez-Rodriguez, Enrique and Beck, Rainer and Clark, Susan E. and Ntormousi, Evangelia and Tassis, Konstantinos and Martin-Alvarez, Sergio and Tahani, Mehrnoosh and Dale, Daniel A. and Castro, Ignacio del Moral and Roman-Duval, Julia and Marcum, Pamela M. and Beckman, John E. and Subramanian, Kandaswamy and Eftekharzadeh, Sarah and Proudfit, Leslie},
	language = {en},
	month = jun,
	note = {arXiv:2303.13586 [astro-ph]},
	publisher = {arXiv},
	shorttitle = {Extragalactic magnetism with {SOFIA} ({SALSA} {Legacy} {Program}) -- {V}},
	title = {Extragalactic magnetism with {SOFIA} ({SALSA} {Legacy} {Program}) -- {V}: {First} results on the magnetic field orientation of galaxies},
	url = {http://arxiv.org/abs/2303.13586},
	urldate = {2023-06-26},
	year = {2023}}

@article{lee_feathering_2014,
	adsurl = {https://ui.adsabs.harvard.edu/abs/2014ApJ...792..122L},
	archiveprefix = {arXiv},
	author = {{Lee}, Wing-Kit},
	doi = {10.1088/0004-637X/792/2/122},
	eid = {122},
	eprint = {1407.5215},
	journal = {\apj},
	month = sep,
	number = {2},
	pages = {122},
	primaryclass = {astro-ph.GA},
	title = {{Feathering Instability of Spiral Arms. II. Parameter Study}},
	volume = {792},
	year = 2014}

@article{kim_formation_2006,
	adsurl = {https://ui.adsabs.harvard.edu/abs/2006ApJ...646..213K},
	archiveprefix = {arXiv},
	author = {{Kim}, Woong-Tae and {Ostriker}, Eve C.},
	doi = {10.1086/504677},
	eprint = {astro-ph/0603751},
	journal = {\apj},
	month = jul,
	number = {1},
	pages = {213-231},
	primaryclass = {astro-ph},
	title = {{Formation of Spiral-Arm Spurs and Bound Clouds in Vertically Stratified Galactic Gas Disks}},
	volume = {646},
	year = 2006}

@article{khoperskov_global_2018,
	author = {Khoperskov, Sergey A. and Khrapov, Sergey S.},
	doi = {10.1051/0004-6361/201629988},
	issn = {0004-6361, 1432-0746},
	journal = {Astronomy \& Astrophysics},
	language = {en},
	month = jan,
	pages = {A104},
	title = {Global enhancement and structure formation of the magnetic field in spiral galaxies},
	url = {https://www.aanda.org/10.1051/0004-6361/201629988},
	urldate = {2023-08-24},
	volume = {609},
	year = {2018}}

@article{pakmor_magnetic_2017,
	author = {Pakmor, R{\"u}diger and G{\'o}mez, Facundo A. and Grand, Robert J. J. and Marinacci, Federico and Simpson, Christine M. and Springel, Volker and Campbell, David J. R. and Frenk, Carlos S. and Guillet, Thomas and Pfrommer, Christoph and White, Simon D. M.},
	doi = {10.1093/mnras/stx1074},
	issn = {0035-8711, 1365-2966},
	journal = {Monthly Notices of the Royal Astronomical Society},
	language = {en},
	month = aug,
	number = {3},
	pages = {3185--3199},
	title = {Magnetic field formation in the {Milky} {Way} like disc galaxies of the {Auriga} project},
	url = {https://academic.oup.com/mnras/article/469/3/3185/3798210},
	urldate = {2023-08-14},
	volume = {469},
	year = {2017}}

@article{goldbaum_mass_2015,
	author = {Goldbaum, Nathan J. and Krumholz, Mark R. and Forbes, John C.},
	doi = {10.1088/0004-637X/814/2/131},
	issn = {1538-4357},
	journal = {The Astrophysical Journal},
	language = {en},
	month = nov,
	note = {arXiv:1510.08458 [astro-ph]},
	number = {2},
	pages = {131},
	shorttitle = {Mass {Transport} and {Turbulence} in {Gravitationally} {Unstable} {Disk} {Galaxies}. {I}},
	title = {Mass {Transport} and {Turbulence} in {Gravitationally} {Unstable} {Disk} {Galaxies}. {I}: {The} {Case} of {Pure} {Self}-{Gravity}},
	url = {http://arxiv.org/abs/1510.08458},
	urldate = {2023-05-10},
	volume = {814},
	year = {2015}}

@article{vazquezsemadeni_molecular_2007,
	author = {Vazquez‐Semadeni, Enrique and Gomez, Gilberto C. and Jappsen, A. Katharina and Ballesteros‐Paredes, Javier and Gonzalez, Ricardo F. and Klessen, Ralf S.},
	doi = {10.1086/510771},
	issn = {0004-637X, 1538-4357},
	journal = {The Astrophysical Journal},
	language = {en},
	month = mar,
	number = {2},
	pages = {870--883},
	title = {Molecular {Cloud} {Evolution}. {II}. {From} {Cloud} {Formation} to the {Early} {Stages} of {Star} {Formation} in {Decaying} {Conditions}},
	url = {https://iopscience.iop.org/article/10.1086/510771},
	urldate = {2023-08-16},
	volume = {657},
	year = {2007}}

@article{meidt_phangsjwst_2023,
	author = {Meidt, Sharon E. and Rosolowsky, Erik and Sun, Jiayi and Koch, Eric W. and Klessen, Ralf S. and Leroy, Adam K. and Schinnerer, Eva and Barnes, Ashley. T. and Glover, Simon C. O. and Lee, Janice C. and Van Der Wel, Arjen and Watkins, Elizabeth J. and Williams, Thomas G. and Bigiel, F. and Boquien, M{\'e}d{\'e}ric and Blanc, Guillermo A. and Cao, Yixian and Chevance, M{\'e}lanie and Dale, Daniel A. and Egorov, Oleg V. and Emsellem, Eric and Grasha, Kathryn and Henshaw, Jonathan D. and Kruijssen, J. M. Diederik and Larson, Kirsten L. and Liu, Daizhong and Murphy, Eric J. and Pety, J{\'e}r{\^o}me and Querejeta, Miguel and Saito, Toshiki and Sandstrom, Karin M. and Smith, Rowan J. and Sormani, Mattia C. and Thilker, David A.},
	doi = {10.3847/2041-8213/acaaa8},
	issn = {2041-8205, 2041-8213},
	journal = {The Astrophysical Journal Letters},
	language = {en},
	month = feb,
	number = {2},
	pages = {L18},
	shorttitle = {{PHANGS}--{JWST} {First} {Results}},
	title = {{PHANGS}--{JWST} {First} {Results}: {Interstellar} {Medium} {Structure} on the {Turbulent} {Jeans} {Scale} in {Four} {Disk} {Galaxies} {Observed} by {JWST} and the {Atacama} {Large} {Millimeter}/submillimeter {Array}},
	url = {https://iopscience.iop.org/article/10.3847/2041-8213/acaaa8},
	urldate = {2023-08-23},
	volume = {944},
	year = {2023}}

@article{crutcher_review_2019,
	author = {Crutcher, Richard M. and Kemball, Athol J.},
	issn = {2296-987X},
	journal = {Frontiers in Astronomy and Space Sciences},
	title = {Review of {Zeeman} {Effect} {Observations} of {Regions} of {Star} {Formation}},
	url = {https://www.frontiersin.org/articles/10.3389/fspas.2019.00066},
	urldate = {2023-04-25},
	volume = {6},
	year = {2019}}

@article{pakmor_simulations_2013,
	author = {Pakmor, R{\"u}diger and Springel, Volker},
	doi = {10.1093/mnras/stt428},
	issn = {0035-8711},
	journal = {Monthly Notices of the Royal Astronomical Society},
	month = jun,
	number = {1},
	pages = {176--193},
	title = {Simulations of magnetic fields in isolated disc galaxies},
	url = {https://doi.org/10.1093/mnras/stt428},
	urldate = {2023-11-10},
	volume = {432},
	year = {2013}}

@article{kim_threedimensional_2002,
	author = {Kim, Woong‐Tae and Ostriker, Eve C. and Stone, James M.},
	doi = {10.1086/344367},
	issn = {0004-637X, 1538-4357},
	journal = {The Astrophysical Journal},
	language = {en},
	month = dec,
	number = {2},
	pages = {1080--1100},
	title = {Three‐dimensional {Simulations} of {Parker}, {Magneto}‐{Jeans}, and {Swing} {Instabilities} in {Shearing} {Galactic} {Gas} {Disks}},
	url = {https://iopscience.iop.org/article/10.1086/344367},
	urldate = {2023-06-29},
	volume = {581},
	year = {2002}}

@article{fensch_universal_2023,
	author = {Fensch, J{\'e}r{\'e}my and Bournaud, Fr{\'e}d{\'e}ric and Brucy, No{\'e} and Dubois, Yohan and Hennebelle, Patrick and Rosdahl, Joakim},
	doi = {10.1051/0004-6361/202245491},
	issn = {0004-6361, 1432-0746},
	journal = {Astronomy \& Astrophysics},
	language = {en},
	month = apr,
	pages = {A193},
	title = {Universal gravity-driven isothermal turbulence cascade in disk galaxies},
	url = {https://www.aanda.org/10.1051/0004-6361/202245491},
	urldate = {2023-11-14},
	volume = {672},
	year = {2023}}

@article{romeo_wiegert_2011_q_stability,
	adsurl = {https://ui.adsabs.harvard.edu/abs/2011MNRAS.416.1191R},
	archiveprefix = {arXiv},
	author = {{Romeo}, Alessandro B. and {Wiegert}, Joachim},
	doi = {10.1111/j.1365-2966.2011.19120.x},
	eprint = {1101.4519},
	journal = {\mnras},
	month = sep,
	number = {2},
	pages = {1191-1196},
	primaryclass = {astro-ph.CO},
	title = {{The effective stability parameter for two-component galactic discs: is Q$^{-1}$ {\ensuremath{\approx}} Q$^{-1}$$_{stars}$ + Q$^{-1}$$_{gas}$?}},
	volume = {416},
	year = 2011}

@article{robinson_wadsley_2024,
	adsurl = {https://ui.adsabs.harvard.edu/abs/2024MNRAS.534.1420R},
	archiveprefix = {arXiv},
	author = {{Robinson}, Hector and {Wadsley}, James},
	doi = {10.1093/mnras/stae2132},
	eprint = {2310.15244},
	journal = {\mnras},
	month = oct,
	number = {2},
	pages = {1420-1432},
	primaryclass = {astro-ph.GA},
	title = {{Regulating star formation in a magnetized disc galaxy}},
	volume = {534},
	year = 2024}

@article{naomi_2023_hi,
	adsurl = {https://ui.adsabs.harvard.edu/abs/2023ARA&A..61...19M},
	archiveprefix = {arXiv},
	author = {{McClure-Griffiths}, Naomi M. and {Stanimirovi{\'c}}, Sne{\v{z}}ana and {Rybarczyk}, Daniel R.},
	doi = {10.1146/annurev-astro-052920-104851},
	eprint = {2307.08464},
	journal = {\araa},
	month = aug,
	pages = {19-63},
	primaryclass = {astro-ph.GA},
	title = {{Atomic Hydrogen in the Milky Way: A Stepping Stone in the Evolution of Galaxies}},
	volume = {61},
	year = 2023}

@ARTICLE{bland-hawthorn_2023,
       author = {{Bland-Hawthorn}, Joss and {Tepper-Garcia}, Thor and {Agertz}, Oscar and {Freeman}, Ken},
        title = "{The Rapid Onset of Stellar Bars in the Baryon-dominated Centers of Disk Galaxies}",
      journal = {\apj},
         year = 2023,
        month = apr,
       volume = {947},
       number = {2},
          eid = {80},
        pages = {80},
          doi = {10.3847/1538-4357/acc469},
archivePrefix = {arXiv},
       eprint = {2303.05574},
 primaryClass = {astro-ph.GA},
       adsurl = {https://ui.adsabs.harvard.edu/abs/2023ApJ...947...80B}
}

@ARTICLE{bland-hawthorn_2024,
       author = {{Bland-Hawthorn}, Joss and {Tepper-Garcia}, Thor and {Agertz}, Oscar and {Federrath}, Christoph},
        title = "{Turbulent Gas-rich Disks at High Redshift: Bars and Bulges in a Radial Shear Flow}",
      journal = {\apj},
         year = 2024,
        month = jun,
       volume = {968},
       number = {2},
          eid = {86},
        pages = {86},
          doi = {10.3847/1538-4357/ad4118},
archivePrefix = {arXiv},
       eprint = {2402.06060},
 primaryClass = {astro-ph.GA},
       adsurl = {https://ui.adsabs.harvard.edu/abs/2024ApJ...968...86B}
}

@ARTICLE{arora_2025,
       author = {{Arora}, Raghav and {Federrath}, Christoph and {Krumholz}, Mark and {Banerjee}, Robi},
        title = "{Formation of filaments and feathers in disc galaxies: Is self-gravity enough?}",
      journal = {\aap},
         year = 2025,
        month = mar,
       volume = {695},
          eid = {A155},
        pages = {A155},
          doi = {10.1051/0004-6361/202453501},
archivePrefix = {arXiv},
       eprint = {2502.18565},
 primaryClass = {astro-ph.GA},
       adsurl = {https://ui.adsabs.harvard.edu/abs/2025A&A...695A.155A}
}

@ARTICLE{gammie_1996,
       author = {{Gammie}, Charles F.},
        title = "{Linear Theory of Magnetized, Viscous, Self-gravitating Gas Disks}",
      journal = {\apj},
         year = 1996,
        month = may,
       volume = {462},
        pages = {725},
          doi = {10.1086/177185},
       adsurl = {https://ui.adsabs.harvard.edu/abs/1996ApJ...462..725G}
}

@ARTICLE{elmegreen_1987_magnetic,
       author = {{Elmegreen}, Bruce G.},
        title = "{Supercloud Formation by Nonaxisymmetric Gravitational Instabilities in Sheared Magnetic Galaxy Disks}",
      journal = {\apj},
         year = 1987,
        month = jan,
       volume = {312},
        pages = {626},
          doi = {10.1086/164907},
       adsurl = {https://ui.adsabs.harvard.edu/abs/1987ApJ...312..626E}
}

@ARTICLE{kim_ostriker_2001,
       author = {{Kim}, Woong-Tae and {Ostriker}, Eve C.},
        title = "{Amplification, Saturation, and Q Thresholds for Runaway: Growth of Self-Gravitating Structures in Models of Magnetized Galactic Gas Disks}",
      journal = {\apj},
         year = 2001,
        month = sep,
       volume = {559},
       number = {1},
        pages = {70-95},
          doi = {10.1086/322330},
archivePrefix = {arXiv},
       eprint = {astro-ph/0105375},
 primaryClass = {astro-ph},
       adsurl = {https://ui.adsabs.harvard.edu/abs/2001ApJ...559...70K}
}

@INPROCEEDINGS{pattel_ppvii_2023,
       author = {{Pattle}, K. and {Fissel}, L. and {Tahani}, M. and {Liu}, T. and {Ntormousi}, E.},
        title = "{Magnetic Fields in Star Formation: from Clouds to Cores}",
    booktitle = {Protostars and Planets VII},
         year = 2023,
       editor = {{Inutsuka}, S. and {Aikawa}, Y. and {Muto}, T. and {Tomida}, K. and {Tamura}, M.},
       series = {Astronomical Society of the Pacific Conference Series},
       volume = {534},
        month = jul,
        pages = {193},
          doi = {10.48550/arXiv.2203.11179},
archivePrefix = {arXiv},
       eprint = {2203.11179},
 primaryClass = {astro-ph.GA},
       adsurl = {https://ui.adsabs.harvard.edu/abs/2023ASPC..534..193P}
}

@ARTICLE{whitworth_2025,
       author = {{Whitworth}, D.~J. and {Srinivasan}, S. and {Pudritz}, R.~E. and {Mac Low}, M.-M. and {Eadie}, G. and {Palau}, A. and {Soler}, J.~D. and {Smith}, R.~J. and {Pattle}, K. and {Robinson}, H. and et al.},
        title = "{On the relation between magnetic field strength and gas density in the interstellar medium: a multiscale analysis}",
      journal = {\mnras},
         year = 2025,
        month = jul,
       volume = {540},
       number = {3},
        pages = {2762-2786},
          doi = {10.1093/mnras/staf901},
archivePrefix = {arXiv},
       eprint = {2407.18293},
 primaryClass = {astro-ph.GA},
       adsurl = {https://ui.adsabs.harvard.edu/abs/2025MNRAS.540.2762W}
}

@software{federrath_2022_turbGen,
       author = {{Federrath}, C. and {Roman-Duval}, J. and {Klessen}, R.~S. and {Schmidt}, W. and {Mac Low}, M.-M.},
        title = "{TG: Turbulence Generator}",
 howpublished = {Astrophysics Source Code Library, record ascl:2204.001},
         year = 2022,
        month = apr,
          eid = {ascl:2204.001},
archivePrefix = {ascl},
       eprint = {2204.001},
       adsurl = {https://ui.adsabs.harvard.edu/abs/2022ascl.soft04001F}
}

@ARTICLE{lou_zou_2006,
       author = {{Lou}, Yu-Qing and {Zou}, Yue},
        title = "{Axisymmetric stability criteria for a composite system of stellar and magnetized gaseous singular isothermal discs}",
      journal = {\mnras},
         year = 2006,
        month = mar,
       volume = {366},
       number = {3},
        pages = {1037-1049},
          doi = {10.1111/j.1365-2966.2005.09878.x},
archivePrefix = {arXiv},
       eprint = {astro-ph/0511348},
 primaryClass = {astro-ph},
       adsurl = {https://ui.adsabs.harvard.edu/abs/2006MNRAS.366.1037L}
}

@ARTICLE{borlaff_salsaV,
       author = {{Borlaff}, Alejandro S. and {Lopez-Rodriguez}, Enrique and {Beck}, Rainer and {Clark}, Susan E. and {Ntormousi}, Evangelia and {Tassis}, Konstantinos and {Martin-Alvarez}, Sergio and {Tahani}, Mehrnoosh and {Dale}, Daniel A. and {del Moral-Castro}, Ignacio and et al.},
        title = "{Extragalactic Magnetism with SOFIA (SALSA Legacy Program). V. First Results on the Magnetic Field Orientation of Galaxies}",
      journal = {\apj},
         year = 2023,
        month = jul,
       volume = {952},
       number = {1},
          eid = {4},
        pages = {4},
          doi = {10.3847/1538-4357/acd934},
archivePrefix = {arXiv},
       eprint = {2303.13586},
 primaryClass = {astro-ph.GA},
       adsurl = {https://ui.adsabs.harvard.edu/abs/2023ApJ...952....4B}
}

@ARTICLE{giessuebel_beck_2014,
       author = {{Gie{\ss}{\"u}bel}, R. and {Beck}, R.},
        title = "{The magnetic field structure of the central region in M 31}",
      journal = {\aap},
         year = 2014,
        month = nov,
       volume = {571},
          eid = {A61},
        pages = {A61},
          doi = {10.1051/0004-6361/201323211},
archivePrefix = {arXiv},
       eprint = {1408.4582},
 primaryClass = {astro-ph.GA},
       adsurl = {https://ui.adsabs.harvard.edu/abs/2014A&A...571A..61G}
}

@ARTICLE{han_beck_1999,
       author = {{Han}, J.~L. and {Beck}, R. and {Ehle}, M. and {Haynes}, R.~F. and {Wielebinski}, R.},
        title = "{Magnetic fields in the spiral galaxy NGC 2997}",
      journal = {\aap},
         year = 1999,
        month = aug,
       volume = {348},
        pages = {405-417},
       adsurl = {https://ui.adsabs.harvard.edu/abs/1999A&A...348..405H}
}

@ARTICLE{beck_ic342_2015,
       author = {{Beck}, Rainer},
        title = "{Magnetic fields in the nearby spiral galaxy IC 342: A multi-frequency radio polarization study}",
      journal = {\aap},
         year = 2015,
        month = jun,
       volume = {578},
          eid = {A93},
        pages = {A93},
          doi = {10.1051/0004-6361/201425572},
archivePrefix = {arXiv},
       eprint = {1502.05439},
 primaryClass = {astro-ph.GA},
       adsurl = {https://ui.adsabs.harvard.edu/abs/2015A&A...578A..93B}
}

@ARTICLE{beck_2007_magneticArms_ngc6946,
       author = {{Beck}, R.},
        title = "{Magnetism in the spiral galaxy NGC 6946: magnetic arms, depolarization rings, dynamo modes, and helical fields}",
      journal = {\aap},
         year = 2007,
        month = aug,
       volume = {470},
       number = {2},
        pages = {539-556},
          doi = {10.1051/0004-6361:20066988},
archivePrefix = {arXiv},
       eprint = {0705.4163},
 primaryClass = {astro-ph},
       adsurl = {https://ui.adsabs.harvard.edu/abs/2007A&A...470..539B}
}

@ARTICLE{dobbs_pettitt_magneticReversals_2016,
       author = {{Dobbs}, C.~L. and {Price}, D.~J. and {Pettitt}, A.~R. and {Bate}, M.~R. and {Tricco}, T.~S.},
        title = "{Magnetic field evolution and reversals in spiral galaxies}",
      journal = {\mnras},
         year = 2016,
        month = oct,
       volume = {461},
       number = {4},
        pages = {4482-4495},
          doi = {10.1093/mnras/stw1625},
archivePrefix = {arXiv},
       eprint = {1607.05532},
 primaryClass = {astro-ph.GA},
       adsurl = {https://ui.adsabs.harvard.edu/abs/2016MNRAS.461.4482D}
}

@ARTICLE{ryan_rowan_whitworth_2026,
       author = {{McGuiness}, Ryan and {Smith}, Rowan J. and {Whitworth}, David},
        title = "{A bottleneck for star formation: the importance of magnetic fields during the formation of cold gas in galaxies.}",
      journal = {\mnras},
         year = 2026,
        month = jan,
          doi = {10.1093/mnras/stag099},
archivePrefix = {arXiv},
       eprint = {2512.04184},
 primaryClass = {astro-ph.GA},
       adsurl = {https://ui.adsabs.harvard.edu/abs/2026MNRAS.tmp...83M}
}

@ARTICLE{whitworth_smith_2023,
       author = {{Whitworth}, David J. and {Smith}, Rowan J. and {Klessen}, Ralf S. and {Mac Low}, Mordecai-Mark and {Glover}, Simon C.~O. and {Tress}, Robin and {Pakmor}, R{\"u}diger and {Soler}, Juan D.},
        title = "{Magnetic fields do not suppress global star formation in low metallicity dwarf galaxies}",
      journal = {\mnras},
         year = 2023,
        month = mar,
       volume = {520},
       number = {1},
        pages = {89-106},
          doi = {10.1093/mnras/stad105},
archivePrefix = {arXiv},
       eprint = {2210.04922},
 primaryClass = {astro-ph.GA},
       adsurl = {https://ui.adsabs.harvard.edu/abs/2023MNRAS.520...89W}
}

@ARTICLE{wibking_krumholz_2023,
       author = {{Wibking}, Benjamin D. and {Krumholz}, Mark R.},
        title = "{The global structure of magnetic fields and gas in simulated Milky Way-analogue galaxies}",
      journal = {\mnras},
         year = 2023,
        month = jun,
       volume = {521},
       number = {4},
        pages = {5972-5990},
          doi = {10.1093/mnras/stac2648},
archivePrefix = {arXiv},
       eprint = {2105.04136},
 primaryClass = {astro-ph.GA},
       adsurl = {https://ui.adsabs.harvard.edu/abs/2023MNRAS.521.5972W}
}

@ARTICLE{gurman_steinwandel_2025,
       author = {{Gurman}, Alon and {Steinwandel}, Ulrich P. and {Hu}, Chia-Yu and {Sternberg}, Amiel},
        title = "{The GHOSDT Simulations: I. Magnetic Support in Gas-rich Disks}",
      journal = {\apj},
         year = 2025,
        month = may,
       volume = {984},
       number = {2},
          eid = {142},
        pages = {142},
          doi = {10.3847/1538-4357/adc814},
archivePrefix = {arXiv},
       eprint = {2411.10514},
 primaryClass = {astro-ph.GA},
       adsurl = {https://ui.adsabs.harvard.edu/abs/2025ApJ...984..142G}
}

@ARTICLE{jog_1992_swing,
       author = {{Jog}, Chanda J.},
        title = "{Swing Amplification of Nonaxisymmetric Perturbations in Stars and Gas in a Sheared Galactic Disk}",
      journal = {\apj},
         year = 1992,
        month = may,
       volume = {390},
        pages = {378},
          doi = {10.1086/171289},
       adsurl = {https://ui.adsabs.harvard.edu/abs/1992ApJ...390..378J}
}

@ARTICLE{romeo_1992,
       author = {{Romeo}, Alessandro B.},
        title = "{Stability of thick two-component galactic discs}",
      journal = {\mnras},
         year = 1992,
        month = may,
       volume = {256},
       number = {2},
        pages = {307-320},
          doi = {10.1093/mnras/256.2.307},
       adsurl = {https://ui.adsabs.harvard.edu/abs/1992MNRAS.256..307R}
}

@ARTICLE{fan_lou_1997_mhdWavesSwing,
       author = {{Fan}, Zuhui and {Lou}, Yu-Qing},
        title = "{Swing amplification of fast and slow density waves in thin magnetized gaseous discs}",
      journal = {\mnras},
         year = 1997,
        month = oct,
       volume = {291},
       number = {1},
        pages = {91-109},
          doi = {10.1093/mnras/291.1.91},
       adsurl = {https://ui.adsabs.harvard.edu/abs/1997MNRAS.291...91F}
}

@ARTICLE{elmegreen1991,
       author = {{Elmegreen}, Bruce G.},
        title = "{Cloud Formation by Combined Instabilities in Galactic Gas Layers: Evidence for a Q Threshold in the Fragmentation of Shearing Wavelets}",
      journal = {\apj},
         year = 1991,
        month = sep,
       volume = {378},
        pages = {139},
          doi = {10.1086/170414},
       adsurl = {https://ui.adsabs.harvard.edu/abs/1991ApJ...378..139E}
}

@ARTICLE{iffrig_hennebelle_2017,
       author = {{Iffrig}, Olivier and {Hennebelle}, Patrick},
        title = "{Structure distribution and turbulence in self-consistently supernova-driven ISM of multiphase magnetized galactic discs}",
      journal = {\aap},
         year = 2017,
        month = aug,
       volume = {604},
          eid = {A70},
        pages = {A70},
          doi = {10.1051/0004-6361/201630290},
archivePrefix = {arXiv},
       eprint = {1703.10421},
 primaryClass = {astro-ph.GA},
       adsurl = {https://ui.adsabs.harvard.edu/abs/2017A&A...604A..70I}
}

@ARTICLE{kim_ostriker_2015,
       author = {{Kim}, Chang-Goo and {Ostriker}, Eve C.},
        title = "{Vertical Equilibrium, Energetics, and Star Formation Rates in Magnetized Galactic Disks Regulated by Momentum Feedback from Supernovae}",
      journal = {\apj},
         year = 2015,
        month = dec,
       volume = {815},
       number = {1},
          eid = {67},
        pages = {67},
          doi = {10.1088/0004-637X/815/1/67},
archivePrefix = {arXiv},
       eprint = {1511.00010},
 primaryClass = {astro-ph.GA},
       adsurl = {https://ui.adsabs.harvard.edu/abs/2015ApJ...815...67K}
}

@ARTICLE{girichidis_2018,
       author = {{Girichidis}, Philipp and {Seifried}, Daniel and {Naab}, Thorsten and {Peters}, Thomas and {Walch}, Stefanie and {W{\"u}nsch}, Richard and {Glover}, Simon C.~O. and {Klessen}, Ralf S.},
        title = "{The SILCC project - V. The impact of magnetic fields on the chemistry and the formation of molecular clouds}",
      journal = {\mnras},
         year = 2018,
        month = nov,
       volume = {480},
       number = {3},
        pages = {3511-3540},
          doi = {10.1093/mnras/sty2016},
archivePrefix = {arXiv},
       eprint = {1808.05222},
 primaryClass = {astro-ph.GA},
       adsurl = {https://ui.adsabs.harvard.edu/abs/2018MNRAS.480.3511G}
}

@ARTICLE{thor_nexus2024,
       author = {{Tepper-Garc{\'\i}a}, Thor and {Bland-Hawthorn}, Joss and {Vasiliev}, Eugene and {Agertz}, Oscar and {Teyssier}, Romain and {Federrath}, Christoph},
        title = "{NEXUS: a framework for controlled simulations of idealized galaxies}",
      journal = {\mnras},
         year = 2024,
        month = nov,
       volume = {535},
       number = {1},
        pages = {187-206},
          doi = {10.1093/mnras/stae2372},
archivePrefix = {arXiv},
       eprint = {2406.00342},
 primaryClass = {astro-ph.GA},
       adsurl = {https://ui.adsabs.harvard.edu/abs/2024MNRAS.535..187T}
}

@ARTICLE{han2017_observingMagneticFields,
       author = {{Han}, J.~L.},
        title = "{Observing Interstellar and Intergalactic Magnetic Fields}",
      journal = {\araa},
         year = 2017,
        month = aug,
       volume = {55},
       number = {1},
        pages = {111-157},
          doi = {10.1146/annurev-astro-091916-055221},
       adsurl = {https://ui.adsabs.harvard.edu/abs/2017ARA&A..55..111H}
}

@ARTICLE{bertin1989,
       author = {{Bertin}, G. and {Lin}, C.~C. and {Lowe}, S.~A. and {Thurstans}, R.~P.},
        title = "{Modal Approach to the Morphology of Spiral Galaxies. I. Basic Structure and Astrophysical Viability}",
      journal = {\apj},
         year = 1989,
        month = mar,
       volume = {338},
        pages = {78},
          doi = {10.1086/167182},
       adsurl = {https://ui.adsabs.harvard.edu/abs/1989ApJ...338...78B}
}

@ARTICLE{renaudRomeoAgertz_2021,
       author = {{Renaud}, Florent and {Romeo}, Alessandro B. and {Agertz}, Oscar},
        title = "{From giant clumps to clouds - I. The impact of gas fraction evolution on the stability of galactic discs}",
      journal = {\mnras},
         year = 2021,
        month = nov,
       volume = {508},
       number = {1},
        pages = {352-370},
          doi = {10.1093/mnras/stab2604},
archivePrefix = {arXiv},
       eprint = {2106.00020},
 primaryClass = {astro-ph.GA},
       adsurl = {https://ui.adsabs.harvard.edu/abs/2021MNRAS.508..352R}
}

@ARTICLE{sellwood1980_bars,
       author = {{Sellwood}, J.~A.},
        title = "{Galaxy models with live halos}",
      journal = {\aap},
         year = 1980,
        month = sep,
       volume = {89},
       number = {3},
        pages = {296-307},
       adsurl = {https://ui.adsabs.harvard.edu/abs/1980A&A....89..296S}
}

@ARTICLE{sellwood2014_bar,
       author = {{Sellwood}, J.~A.},
        title = "{Secular evolution in disk galaxies}",
      journal = {Reviews of Modern Physics},
         year = 2014,
        month = jan,
       volume = {86},
       number = {1},
        pages = {1-46},
          doi = {10.1103/RevModPhys.86.1},
archivePrefix = {arXiv},
       eprint = {1310.0403},
 primaryClass = {astro-ph.GA},
       adsurl = {https://ui.adsabs.harvard.edu/abs/2014RvMP...86....1S}
}

@ARTICLE{mandelkarDekel2014,
       author = {{Mandelker}, Nir and {Dekel}, Avishai and {Ceverino}, Daniel and {Tweed}, Dylan and {Moody}, Christopher E. and {Primack}, Joel},
        title = "{The population of giant clumps in simulated high-z galaxies: in situ and ex situ migration and survival}",
      journal = {\mnras},
         year = 2014,
        month = oct,
       volume = {443},
       number = {4},
        pages = {3675-3702},
          doi = {10.1093/mnras/stu1340},
archivePrefix = {arXiv},
       eprint = {1311.0013},
 primaryClass = {astro-ph.CO},
       adsurl = {https://ui.adsabs.harvard.edu/abs/2014MNRAS.443.3675M}
}

@ARTICLE{dekelCeverino2009_clumps,
       author = {{Dekel}, Avishai and {Sari}, Re'em and {Ceverino}, Daniel},
        title = "{Formation of Massive Galaxies at High Redshift: Cold Streams, Clumpy Disks, and Compact Spheroids}",
      journal = {\apj},
         year = 2009,
        month = sep,
       volume = {703},
       number = {1},
        pages = {785-801},
          doi = {10.1088/0004-637X/703/1/785},
archivePrefix = {arXiv},
       eprint = {0901.2458},
 primaryClass = {astro-ph.GA},
       adsurl = {https://ui.adsabs.harvard.edu/abs/2009ApJ...703..785D}
}

@ARTICLE{krumholzBurkhart2018,
       author = {{Krumholz}, Mark R. and {Burkhart}, Blakesley and {Forbes}, John C. and {Crocker}, Roland M.},
        title = "{A unified model for galactic discs: star formation, turbulence driving, and mass transport}",
      journal = {\mnras},
         year = 2018,
        month = jun,
       volume = {477},
       number = {2},
        pages = {2716-2740},
          doi = {10.1093/mnras/sty852},
archivePrefix = {arXiv},
       eprint = {1706.00106},
 primaryClass = {astro-ph.GA},
       adsurl = {https://ui.adsabs.harvard.edu/abs/2018MNRAS.477.2716K}
}

@ARTICLE{ferriere2001_reviewISM,
       author = {{Ferri{\`e}re}, Katia M.},
        title = "{The interstellar environment of our galaxy}",
      journal = {Reviews of Modern Physics},
         year = 2001,
        month = oct,
       volume = {73},
       number = {4},
        pages = {1031-1066},
          doi = {10.1103/RevModPhys.73.1031},
archivePrefix = {arXiv},
       eprint = {astro-ph/0106359},
 primaryClass = {astro-ph},
       adsurl = {https://ui.adsabs.harvard.edu/abs/2001RvMP...73.1031F}
}

@ARTICLE{ntormousi2020_galacticDynamo,
       author = {{Ntormousi}, Evangelia and {Tassis}, Konstantinos and {Del Sordo}, Fabio and {Fragkoudi}, Francesca and {Pakmor}, R{\"u}diger},
        title = "{A dynamo amplifying the magnetic field of a Milky-Way-like galaxy}",
      journal = {\aap},
         year = 2020,
        month = sep,
       volume = {641},
          eid = {A165},
        pages = {A165},
          doi = {10.1051/0004-6361/202037835},
archivePrefix = {arXiv},
       eprint = {2006.12574},
 primaryClass = {astro-ph.GA},
       adsurl = {https://ui.adsabs.harvard.edu/abs/2020A&A...641A.165N}
}

@ARTICLE{pakmorRebekka2024,
       author = {{Pakmor}, R{\"u}diger and {Bieri}, Rebekka and {van de Voort}, Freeke and {Werhahn}, Maria and {Fattahi}, Azadeh and {Guillet}, Thomas and {Pfrommer}, Christoph and {Springel}, Volker and {Talbot}, Rosie Y.},
        title = "{Magnetic field amplification in cosmological zoom simulations from dwarf galaxies to galaxy groups}",
      journal = {\mnras},
         year = 2024,
        month = feb,
       volume = {528},
       number = {2},
        pages = {2308-2325},
          doi = {10.1093/mnras/stae112},
archivePrefix = {arXiv},
       eprint = {2309.13104},
 primaryClass = {astro-ph.GA},
       adsurl = {https://ui.adsabs.harvard.edu/abs/2024MNRAS.528.2308P}
}

@ARTICLE{steinwandel2019,
       author = {{Steinwandel}, U.~P. and {Beck}, M.~C. and {Arth}, A. and {Dolag}, K. and {Moster}, B.~P. and {Nielaba}, P.},
        title = "{Magnetic buoyancy in simulated galactic discs with a realistic circumgalactic medium}",
      journal = {\mnras},
         year = 2019,
        month = feb,
       volume = {483},
       number = {1},
        pages = {1008-1028},
          doi = {10.1093/mnras/sty3083},
archivePrefix = {arXiv},
       eprint = {1808.09975},
 primaryClass = {astro-ph.GA},
       adsurl = {https://ui.adsabs.harvard.edu/abs/2019MNRAS.483.1008S}
}

@ARTICLE{bieriPakmor2026,
       author = {{Bieri}, Rebekka and {Pakmor}, R{\"u}diger and {van de Voort}, Freeke and {Talbot}, Rosie Y. and {Werhahn}, Maria and {Pfrommer}, Christoph and {Springel}, Volker},
        title = "{Unveiling the impact of cosmic rays on the disc sizes and outflows from dwarf scales to galaxy groups}",
      journal = {\mnras},
         year = 2026,
        month = apr,
       volume = {547},
       number = {2},
          eid = {stag216},
        pages = {stag216},
          doi = {10.1093/mnras/stag216},
archivePrefix = {arXiv},
       eprint = {2509.07124},
 primaryClass = {astro-ph.GA},
       adsurl = {https://ui.adsabs.harvard.edu/abs/2026MNRAS.547ag216B}
}

@ARTICLE{pillepichAnnalisa2018,
       author = {{Pillepich}, Annalisa and {Springel}, Volker and {Nelson}, Dylan and {Genel}, Shy and {Naiman}, Jill and {Pakmor}, R{\"u}diger and {Hernquist}, Lars and {Torrey}, Paul and {Vogelsberger}, Mark and {Weinberger}, Rainer and et al.},
        title = "{Simulating galaxy formation with the IllustrisTNG model}",
      journal = {\mnras},
         year = 2018,
        month = jan,
       volume = {473},
       number = {3},
        pages = {4077-4106},
          doi = {10.1093/mnras/stx2656},
archivePrefix = {arXiv},
       eprint = {1703.02970},
 primaryClass = {astro-ph.GA},
       adsurl = {https://ui.adsabs.harvard.edu/abs/2018MNRAS.473.4077P}
}

@ARTICLE{hopkinsFIRE2018,
       author = {{Hopkins}, Philip F. and {Wetzel}, Andrew and {Kere{\v{s}}}, Du{\v{s}}an and {Faucher-Gigu{\`e}re}, Claude-Andr{\'e} and {Quataert}, Eliot and {Boylan-Kolchin}, Michael and {Murray}, Norman and {Hayward}, Christopher C. and {Garrison-Kimmel}, Shea and {Hummels}, Cameron and {Feldmann}, Robert and {Torrey}, Paul and {Ma}, Xiangcheng and {Angl{\'e}s-Alc{\'a}zar}, Daniel and {Su}, Kung-Yi and {Orr}, Matthew and {Schmitz}, Denise and {Escala}, Ivanna and {Sanderson}, Robyn and {Grudi{\'c}}, Michael Y. and {Hafen}, Zachary and {Kim}, Ji-Hoon and {Fitts}, Alex and {Bullock}, James S. and {Wheeler}, Coral and {Chan}, T.~K. and {Elbert}, Oliver D. and {Narayanan}, Desika},
        title = "{FIRE-2 simulations: physics versus numerics in galaxy formation}",
      journal = {\mnras},
         year = 2018,
        month = oct,
       volume = {480},
       number = {1},
        pages = {800-863},
          doi = {10.1093/mnras/sty1690},
archivePrefix = {arXiv},
       eprint = {1702.06148},
 primaryClass = {astro-ph.GA},
       adsurl = {https://ui.adsabs.harvard.edu/abs/2018MNRAS.480..800H}
}

@ARTICLE{pakmorAurigaMagnetic2017,
       author = {{Pakmor}, R{\"u}diger and {G{\'o}mez}, Facundo A. and {Grand}, Robert J.~J. and {Marinacci}, Federico and {Simpson}, Christine M. and {Springel}, Volker and {Campbell}, David J.~R. and {Frenk}, Carlos S. and {Guillet}, Thomas and {Pfrommer}, Christoph and {White}, Simon D.~M.},
        title = "{Magnetic field formation in the Milky Way like disc galaxies of the Auriga project}",
      journal = {\mnras},
         year = 2017,
        month = aug,
       volume = {469},
       number = {3},
        pages = {3185-3199},
          doi = {10.1093/mnras/stx1074},
archivePrefix = {arXiv},
       eprint = {1701.07028},
 primaryClass = {astro-ph.GA},
       adsurl = {https://ui.adsabs.harvard.edu/abs/2017MNRAS.469.3185P}
}

@ARTICLE{vasilievAGAMA2019,
       author = {{Vasiliev}, Eugene},
        title = "{AGAMA: action-based galaxy modelling architecture}",
      journal = {\mnras},
         year = 2019,
        month = jan,
       volume = {482},
       number = {2},
        pages = {1525-1544},
          doi = {10.1093/mnras/sty2672},
archivePrefix = {arXiv},
       eprint = {1802.08239},
 primaryClass = {astro-ph.GA},
       adsurl = {https://ui.adsabs.harvard.edu/abs/2019MNRAS.482.1525V}
}

@ARTICLE{widrowEquilibriumICs2008,
       author = {{Widrow}, Lawrence M. and {Pym}, Brent and {Dubinski}, John},
        title = "{Dynamical Blueprints for Galaxies}",
      journal = {\apj},
         year = 2008,
        month = jun,
       volume = {679},
       number = {2},
        pages = {1239-1259},
          doi = {10.1086/587636},
archivePrefix = {arXiv},
       eprint = {0801.3414},
 primaryClass = {astro-ph},
       adsurl = {https://ui.adsabs.harvard.edu/abs/2008ApJ...679.1239W}
}

@ARTICLE{sellwoodGalaxyPackage2014,
       author = {{Sellwood}, J.~A.},
        title = "{GALAXY package for N-body simulation}",
      journal = {arXiv e-prints},
         year = 2014,
        month = jun,
          eid = {arXiv:1406.6606},
        pages = {arXiv:1406.6606},
          doi = {10.48550/arXiv.1406.6606},
archivePrefix = {arXiv},
       eprint = {1406.6606},
 primaryClass = {astro-ph.IM},
       adsurl = {https://ui.adsabs.harvard.edu/abs/2014arXiv1406.6606S}
}

@ARTICLE{yurinSpringelGaelic2014,
       author = {{Yurin}, Denis and {Springel}, Volker},
        title = "{An iterative method for the construction of N-body galaxy models in collisionless equilibrium}",
      journal = {\mnras},
         year = 2014,
        month = oct,
       volume = {444},
       number = {1},
        pages = {62-79},
          doi = {10.1093/mnras/stu1421},
archivePrefix = {arXiv},
       eprint = {1402.1623},
 primaryClass = {astro-ph.CO},
       adsurl = {https://ui.adsabs.harvard.edu/abs/2014MNRAS.444...62Y}
}

@ARTICLE{riederTeyssier_ssd_2016,
       author = {{Rieder}, Michael and {Teyssier}, Romain},
        title = "{A small-scale dynamo in feedback-dominated galaxies as the origin of cosmic magnetic fields - I. The kinematic phase}",
      journal = {\mnras},
         year = 2016,
        month = apr,
       volume = {457},
       number = {2},
        pages = {1722-1738},
          doi = {10.1093/mnras/stv2985},
archivePrefix = {arXiv},
       eprint = {1506.00849},
 primaryClass = {astro-ph.GA},
       adsurl = {https://ui.adsabs.harvard.edu/abs/2016MNRAS.457.1722R}
}

@ARTICLE{wissingShen2023,
       author = {{Wissing}, Robert and {Shen}, Sijing},
        title = "{Numerical dependencies of the galactic dynamo in isolated galaxies with SPH}",
      journal = {\aap},
         year = 2023,
        month = may,
       volume = {673},
          eid = {A47},
        pages = {A47},
          doi = {10.1051/0004-6361/202244753},
archivePrefix = {arXiv},
       eprint = {2208.07889},
 primaryClass = {astro-ph.GA},
       adsurl = {https://ui.adsabs.harvard.edu/abs/2023A&A...673A..47W}
}

@ARTICLE{federrath2010_forcingInSimsObs,
       author = {{Federrath}, C. and {Roman-Duval}, J. and {Klessen}, R.~S. and {Schmidt}, W. and {Mac Low}, M.-M.},
        title = "{Comparing the statistics of interstellar turbulence in simulations and observations. Solenoidal versus compressive turbulence forcing}",
      journal = {\aap},
         year = 2010,
        month = mar,
       volume = {512},
          eid = {A81},
        pages = {A81},
          doi = {10.1051/0004-6361/200912437},
archivePrefix = {arXiv},
       eprint = {0905.1060},
 primaryClass = {astro-ph.SR},
       adsurl = {https://ui.adsabs.harvard.edu/abs/2010A&A...512A..81F}
}


\appendix

\section{ Fastest growing mode of the magneto-Jeans instability} \label{appendix:lengthScale_magneto-Jeans}

In this Appendix we show that the approximation $k_{\rm magneto-Jeans} \approx k_{\rm T}$ used in \autoref{subsect:comparisonWithLinearTheory} does not significantly impact the analysis. Here, $k_{\rm magneto-Jeans}$ is the fastest growing mode of the magneto-Jeans instability and $k_{\rm T}$ is the same for the Toomre instability. We estimate the former from the dispersion relation (c.f. \autoref{eqn:dispersionMJI_dimensionfull}), which we re-cast in terms of the dimensionless parameters of the system $(Q, \beta)$ as  
\begin{equation} \label{eqn:mji_dispersion_dimensionless}
    \begin{split}
    \tilde{\gamma} ^{4} +  \left [ 1 - \frac{2}{Q_{\mathrm{g}}^{2}} |\tilde{k}| + \tilde{k}^{2} \left (1 + \frac{2}{\beta} \right ) \frac{1}{\qgas^{2}}  \right] \tilde{\gamma}^{2} \\  +  \frac{2}{\beta \qgas^{4}} \left [ \tilde{k}^{2} - 2 \tilde{k} \right] \tilde{k}_{y}^{2}  = 0,
    \end{split}
\end{equation}
where $\tilde{\gamma} = \gamma/\kappa$ is the dimensionless growth rate and $\tilde{k}_{x} = k_{x}/k_{\rm T}$ (and similarly for $\tilde{k}_y$) is the dimensionless wavenumber normalised by the Toomre wavenumber $k_{\mathrm{T}} = \pi G \Sigma_0/\cs ^{2}$. We define $k_{\rm magneto-Jeans}$ as the value of $k$ for which $\gamma$ reaches its maximum; since the dispersion relation is a fourth-order polynomial we cannot write down a simple closed-form expression for this value, but for specified $\{\qgas(\beta), \beta \}$ it is straightforward to find numerically. We sample $\{\qgas(\beta), \beta \}$ in a range that corresponds to the initial conditions of the simulation set, subject to the constraint that the initial $Q_{\rm eff}$ profile remains constant. This gives 
\begin{equation}
    Q_{\mathrm{g}}= \frac{Q_{\mathrm{eff}} (R) \xi}{\sqrt{ 1 + 1/\beta}}, 
\end{equation}
where $\xi = 1.325$ is a dimensionless parameter for the thickness correction of the simulations (see \autoref{subsect:comparisonWithLinearTheory} for details). We then sample the $\beta$ in the range $\beta \in (0.1, 10)$, which is the $\beta$ range where magnetic fields are dynamically relevant to the instability (see \autoref{fig:growthRate_with_pbeta}). 

We show $\Delta k/k_{\rm T}$ against $\beta$ in \autoref{fig:mji_fastestGrowingMode}. Here, $\Delta k = k_{\rm magneto-Jeans} - k_{\rm T}$ is the difference between the two wavenumbers. We can see that the $\Delta k/k_{\rm T} \approx25\%$ at $\beta = 10$, and that it decreases to $\approx 1\%$ at $\beta = 0.1$. Considering the analytical curves in \autoref{fig:mFil_with_pbeta}, where we use $m_{\mathrm{magneto-Jeans}} =  k_{\rm magneto-Jeans} R \approx k_{\mathrm{T}}R$, we can see that this additional source of error will only affect the upper error bounds for $\beta \sim 10$ by a factor of two, thus not affecting the comparison with simulations since they underestimate $m_{\rm fil}$.

\begin{figure}
    \centering
     \includegraphics[width=0.95\linewidth]{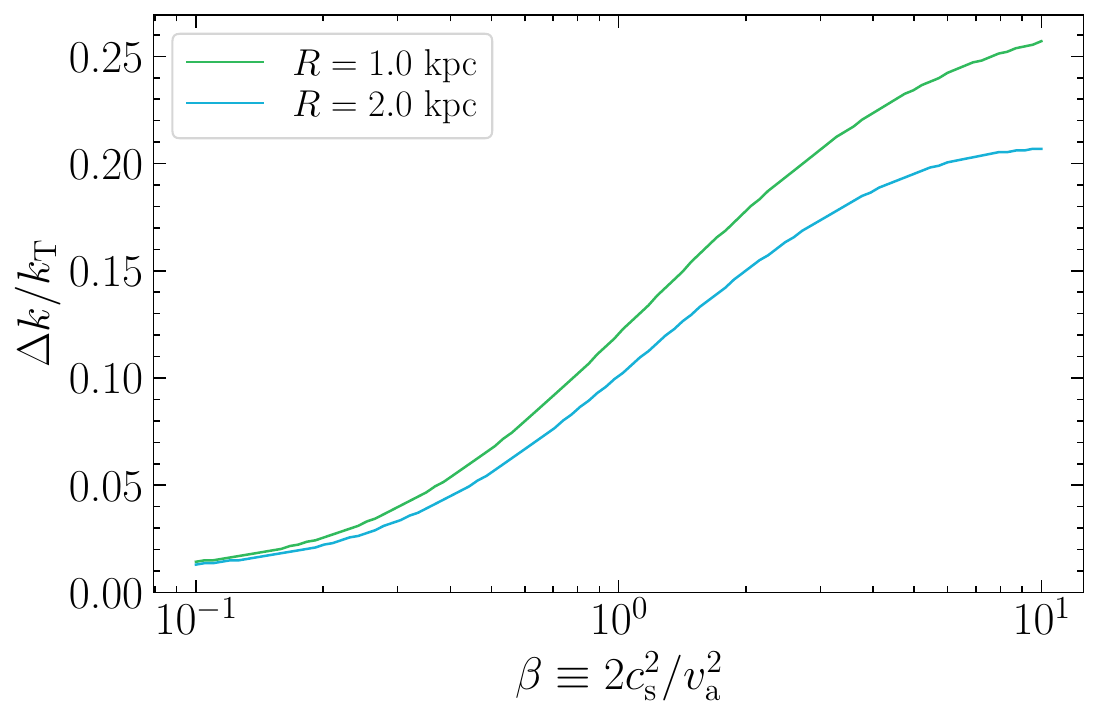}
    \caption{The relative difference between $k_{\rm magneto-Jeans}$ and $k_{\rm T}$, $k_{\rm magneto-Jeans}/k_{\rm T} - 1$, plotted against $\beta$. The different curves show the same quantity calculated at 2 different galactocentric radii, as indicated in the legend. The $k_{\rm magneto-Jeans}$ is calculated using the dispersion relation given by \autoref{eqn:dispersionMJI_dimensionfull} (see text for details), and $k_{\mathrm{T}} = \pi G \Sigma_0/\cs ^{2}$. We do this for model galaxies that share their initial conditions with the simulation set in the main body of the text. We can see that the relative difference between the two declines from $25\%$ at $\beta = 10$ to $\approx 1 \%$ at $\beta = 0.1$.}
    \label{fig:mji_fastestGrowingMode}
\end{figure}

\section{Runs with the same initial $\qgas$} \label{appendix:runsWithSameQg}

Here, we compare the evolution of two simulated galaxies that share their initial $\langle \qgas \rangle =0.94$, but have a varying $\beta \in \{ \infty, 1\}$. This is complimentary to the simulation suite in the main body of the text, where the runs with varying $\beta$ share their initial $Q_{\rm eff}$. The parameters of the two runs are shown in \autoref{tab:initialConditions_QgRuns}.

In \autoref{fig:qgas1Beta1_projection}, we show the projected surface density of the two galaxies at $t \approx 0.8~\trot$. The left panel is for the $\beta = \infty$ case, and the right one is for the $\beta = 1$ run. We see that both the galaxies form dense structures. Moreover, as seen in \autoref{subsect:discMorphology}, the region where dense structures emerge moves inwards in the $\beta = 1$ case. This magnetised run has the region $R\leq 1.5~\rm kpc$ populated with dense clumps, while the $\beta = \infty$ case remains stable. At $R = 2.0~\rm kpc$, both the runs have dens features throughout their azimuth. Further out, at $R = 2.5~\rm kpc$, we see that the hydrodynamical case has denser features when compared with the magnetised one. 

To quantify the radius-dependent difference in their evolution, we plot the time evolution of the surface density in \autoref{fig:qgas1Beta1_clumpingFactorTimeEvol}. Similar to \autoref{fig:timeEvolotion_standardDeviation_relativeSurfaceDensity}, the two lines represent the two runs and different panels are for $100~\rm pc$ wide bins centred at $R = \{1.0, 1.5, 2.0, 2.5\}~\rm kpc$, as indicated in the legend. Similar to the main body of the text (see \autoref{subsect:filamentFormation_timescales}), here we focus on the exponentially-rising part of the curve. This represents dense structure/feather formation. At $R = 1~\rm kpc$, we see that the magnetised run shows exponential rise while the $\beta = \infty$ case does not. At $R = 1.5~\rm kpc$, the $\beta = \infty$ case starts exhibiting growth, but with a growth rate that lags behind the $\beta = 1$ case by $\approx 20~\%$. This difference decreases systematically as we move radially outwards. At $R = 2.0~\rm kpc$, this is $\approx 3~\%$, and further out at $R=2.5~\rm kpc$ the $\beta = \infty$ case overtakes the $\beta = 1$ run and grows at a faster rate. 

Thus, we see that equipartition magnetic fields destabilise by increasing the growth rate of dense structures for the inner regions of our galaxies with $R\leq 2.0 ~\rm kpc$ ($q\leq 0.36$) even when we do not correct for the additional pressure component that arises when we introduce magnetic fields.

\begin{figure*}
    \centering
     \includegraphics[width=0.95\linewidth]{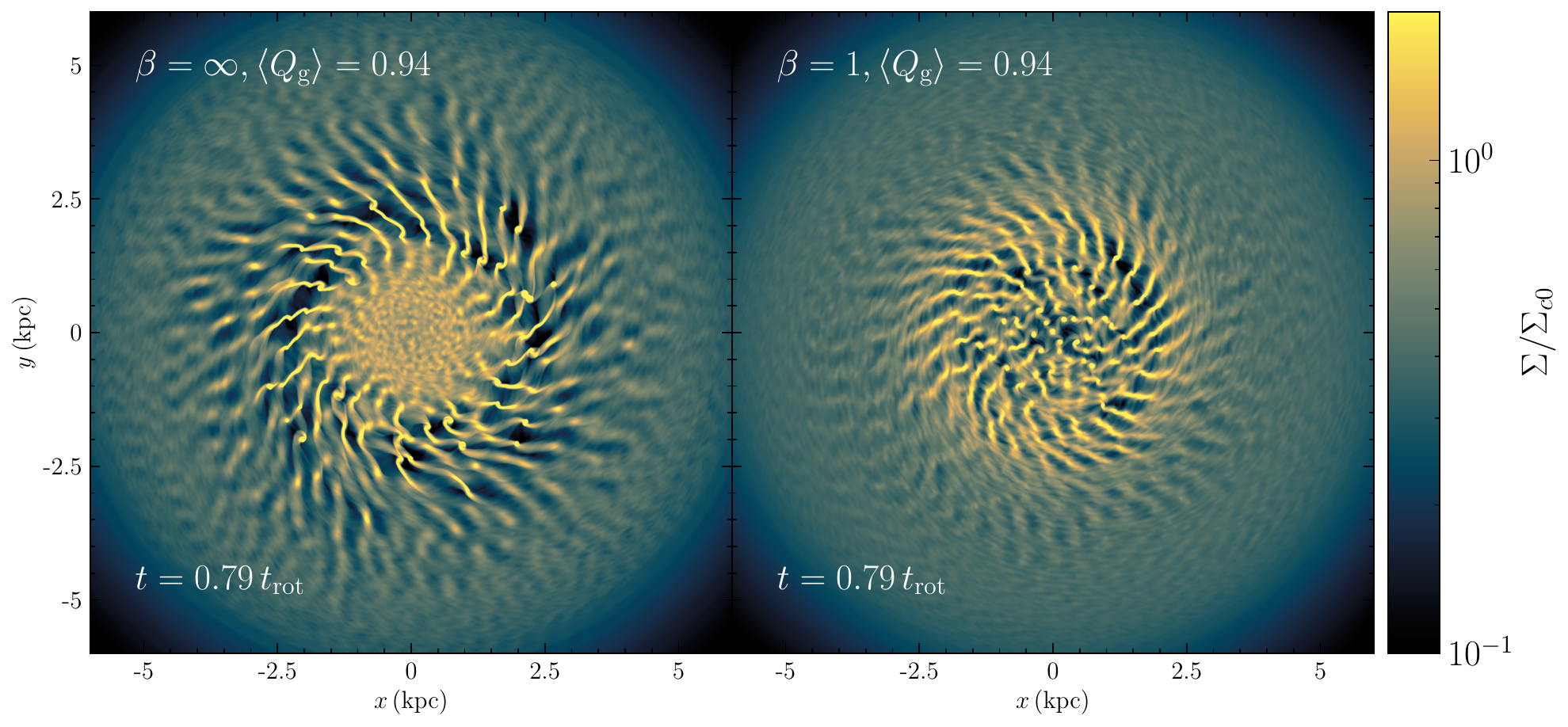}
    \caption{Projected density of the runs with the same initial $\qgas$ profile but $\beta \in \{ \infty, 1\}$ onto the $z$-plane.    The left panel is for the $\beta = \infty$ (hydrodynamical case) and the right panel is for the $\beta =1$ run. Similar to the last row in \autoref{fig:projection_evolution}, both simulations are shown at $t = 0.79~\trot$, where $\trot$ is calculated at $R = 3~\rm kpc$. We can see that both galaxies forms dense structures. The stable region around $R = 1.0~\rm kpc$ in the $\beta = \infty$ case is unstable in the $\beta = 1$ run. At $R = 3.0~\rm kpc$ the $\beta = \infty$ run is relatively more stable.}
    \label{fig:qgas1Beta1_projection}
\end{figure*}

\begin{figure*}
    \centering
     \includegraphics[width=0.95\linewidth]{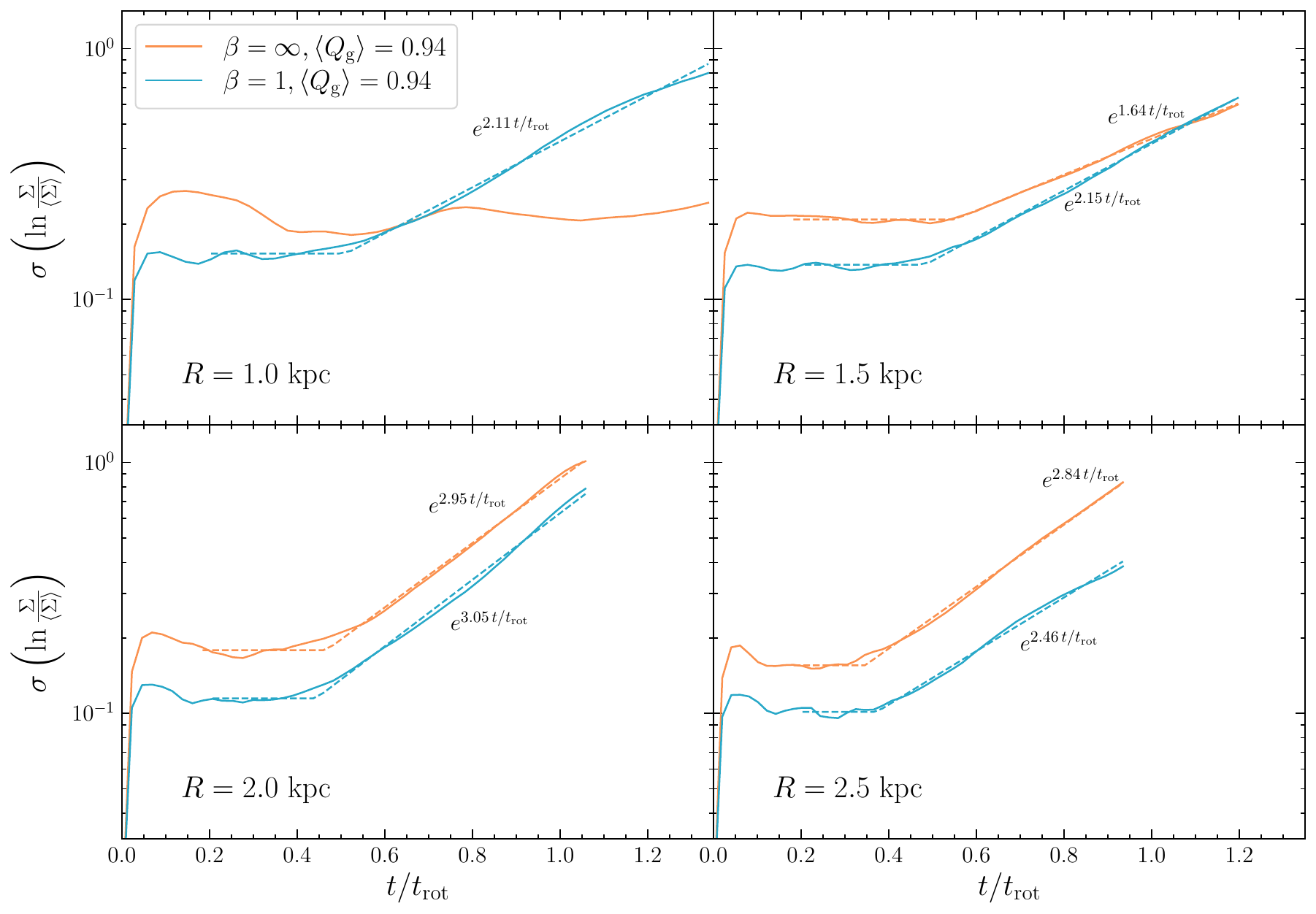}
    \caption{Identical to \autoref{fig:timeEvolotion_standardDeviation_relativeSurfaceDensity}, but for runs with the same $Q_{\rm g}$ profile and with $\beta \in \{ \infty, 1\}$. For $R \leq 2.0 ~\rm kpc$, we see that the $\beta = 1$ run exhibits a steeper exponential growth than the $\beta = \infty$ (hydrodynamical) run. Further out at $R = 2.5~\rm kpc$, this trend is reversed.}
    \label{fig:qgas1Beta1_clumpingFactorTimeEvol}
\end{figure*}

\begin{table}
\setlength{\tabcolsep}{3.2pt} 
\centering
\caption{Similar to \autoref{tab:initialConditions}, but for runs with the same initial $\langle Q_{\rm g} \rangle = 0.94$, $\machc = 28.6$, and varying $\beta$.} 
\begin{tabular}{ccccccc} 
\hline 
\hline
 Model  &$\langle Q_{\rm eff} \rangle $ &$\beta$  & $\Sigma_{\circ}$ & $\langle B \rangle$ \\
  Name & & & $(\rm M_{\odot} \, pc^{-2})$  & $(\mu \rm G)$\\
\hline
$\beta=\infty$ &   0.94    &  $\infty$ &  179 & 0 \\
$Q_{\mathrm{g}}=1, \beta=1$    &   1.33        &  1  &  179 & 13.5 \\
\hline
\hline
\label{tab:initialConditions_QgRuns}
\end{tabular}
\end{table}


\bsp	
\label{lastpage}
\end{document}